\documentclass[11pt, a4paper]{article}

\usepackage{setspace}
\usepackage{bm}
\usepackage{upgreek}
\usepackage{rotating}
\usepackage{graphicx}
\usepackage{verbatim}
\usepackage{wrapfig}
\usepackage[margin=2.5cm]{geometry}
\usepackage{amsmath}
\usepackage{placeins}
\usepackage{url}

\usepackage{cite}

\usepackage{fancyhdr}
\usepackage{color}
\usepackage{amsthm}
\usepackage{amsmath, amssymb}
\usepackage{longtable}
\usepackage{lscape}
\usepackage{multicol}
\usepackage{pdfpages}
\usepackage{siunitx}
\usepackage{graphicx}
\usepackage{enumerate}

\usepackage[affil-it]{authblk}

\usepackage{caption}
\usepackage{subcaption}

\usepackage{bbm}
\usepackage{nameref}

\usepackage{tikz}
\usetikzlibrary{positioning}
\tikzset{>=stealth}
\usetikzlibrary{decorations.markings}
\usetikzlibrary{patterns}
\usetikzlibrary{math}
\usetikzlibrary{calc}

\usepackage{hyperref}

\usepackage{chngcntr}
\counterwithout{table}{section}

\newtheorem{all}{Theorem}[section]

\theoremstyle{plain}
\newtheorem{thm2}[all]{Lemma}
\newtheorem{thm3}[all]{Theorem}

\newtheorem{thm6}[all]{Proposition}

\theoremstyle{definition}
\newtheorem{thm1}[all]{Definition}

\newtheorem{thm5}[all]{Remark}
\newtheorem{thm8}[all]{Assumption}

\newcommand{\BH}{{\mathbb{H}}}
\newcommand{\BI}{{\mathbb{I}}}

\newcommand{\BR}{{\mathbb{R}}}

\newcommand{\BZ}{{\mathbb{Z}}}
\newcommand{\id}{{\mathbb{I}}}

\newcommand{\dd}{{\mathrm{d}}}

\newcommand{\sgn}{\mathrm{sgn}}
\newcommand{\supp}{\mathrm{supp}\, }
\newcommand{\ran}{\mathrm{ran}\ }
\newcommand{\loc}{{\rm loc}}

\newcommand{\com}[1]{}

\newcommand{\R}{\mathbb{R}}

\DeclareMathOperator{\re}{Re}

\makeatletter
\pgfdeclarepatternformonly[\LineSpace]{my north east lines}{\pgfqpoint{-1pt}{-1pt}}{\pgfqpoint{\LineSpace}{\LineSpace}}{\pgfqpoint{\LineSpace}{\LineSpace}}%
{
    \pgfsetcolor{\tikz@pattern@color}
    \pgfsetlinewidth{0.4pt}
    \pgfpathmoveto{\pgfqpoint{0pt}{0pt}}
    \pgfpathlineto{\pgfqpoint{\LineSpace + 0.1pt}{\LineSpace + 0.1pt}}
    \pgfusepath{stroke}
}

\pgfdeclarepatternformonly[\LineSpace]{red north east lines}{\pgfqpoint{-1pt}{-1pt}}{\pgfqpoint{\LineSpace}{\LineSpace}}{\pgfqpoint{\LineSpace}{\LineSpace}}%
{
    \pgfsetcolor{\tikz@pattern@color}
    \pgfsetlinewidth{0.4pt}
    \pgfpathmoveto{\pgfqpoint{0pt}{0pt}}
    \pgfpathlineto{\pgfqpoint{\LineSpace + 0.1pt}{\LineSpace + 0.1pt}}
    \pgfusepath{stroke}
}

\pgfdeclarepatternformonly[\LineSpace]{my north west lines}{\pgfqpoint{-1pt}{-1pt}}{\pgfqpoint{\LineSpace}{\LineSpace}}{\pgfqpoint{\LineSpace}{\LineSpace}}%
{
    \pgfsetcolor{\tikz@pattern@color}
    \pgfsetlinewidth{0.4pt}
    \pgfpathmoveto{\pgfqpoint{0pt}{\LineSpace}}
    \pgfpathlineto{\pgfqpoint{\LineSpace + 0.1pt}{-0.1pt}}
    \pgfusepath{stroke}
}

\pgfdeclarepatternformonly[\LineSpace]{blue north west lines}{\pgfqpoint{-1pt}{-1pt}}{\pgfqpoint{\LineSpace}{\LineSpace}}{\pgfqpoint{\LineSpace}{\LineSpace}}%
{
    \pgfsetcolor{\tikz@pattern@color}
    \pgfsetlinewidth{0.4pt}
    \pgfpathmoveto{\pgfqpoint{0pt}{\LineSpace}}
    \pgfpathlineto{\pgfqpoint{\LineSpace + 0.1pt}{-0.1pt}}
    \pgfusepath{stroke}
}

\pgfdeclarepatternformonly[\LineSpaceP]{red up lines}{\pgfqpoint{-1pt}{-1pt}}{\pgfqpoint{\LineSpaceP}{\LineSpaceP}}{\pgfqpoint{\LineSpaceP}{\LineSpaceP}}%
{
    \pgfsetcolor{\tikz@pattern@color}
    \pgfsetlinewidth{0.4pt}
    \pgfpathmoveto{\pgfqpoint{0pt}{0pt}}
    \pgfpathlineto{\pgfqpoint{0pt}{\LineSpaceP + 0.1pt}}
    \pgfusepath{stroke}
}

\pgfdeclarepatternformonly[\LineSpaceP]{blue up lines}{\pgfqpoint{-1pt}{-1pt}}{\pgfqpoint{\LineSpaceP}{\LineSpaceP}}{\pgfqpoint{\LineSpaceP}{\LineSpaceP}}%
{
    \pgfsetcolor{\tikz@pattern@color}
    \pgfsetlinewidth{0.4pt}
    \pgfpathmoveto{\pgfqpoint{0pt}{0pt}}
    \pgfpathlineto{\pgfqpoint{0pt}{\LineSpaceP + 0.1pt}}
    \pgfusepath{stroke}
}

\makeatother

\newdimen\LineSpace
\tikzset{
    line space/.code={\LineSpace=#1},
    line space=17pt
}

\newdimen\LineSpaceP
\tikzset{
    line space/.code={\LineSpaceP=#1},
    line space=11pt
}

\numberwithin{equation}{section}
\begin{document}
\title{Two-Particle Bound States at Interfaces and Corners}
\author{Barbara Roos\footnote{barbara.roos@ist.ac.at}, Robert Seiringer\footnote{robert.seiringer@ist.ac.at}}
\affil{Institute of Science and Technology Austria,\\ Am Campus 1, 3400 Klosterneuburg, Austria}

\date{\today}

\maketitle

\begin{abstract}
We study two interacting quantum particles forming a bound state in $d$-dimensional free space, and constrain the particles in $k$ directions to $(0,\infty)^k \times \BR^{d-k}$, with Neumann boundary conditions.
First, we prove that the ground state energy strictly decreases upon going from $k$ to $k+1$.
This shows that the particles stick to the corner where all boundary planes intersect.
Second, we show that for all $k$ the resulting Hamiltonian,  after removing the free part of the kinetic energy, has only finitely many eigenvalues below the essential spectrum. 
This paper generalizes the work of Egger, Kerner and Pankrashkin (J. Spectr. Theory 10(4):1413--1444,  2020) to dimensions $d>1$. 
\end{abstract}

\section{Introduction and Main Results}

We consider two interacting quantum particles in $d$-dimensional space that form a bound state in free space.
We constrain the particles in $k$ directions to $(0,\infty)^k \times \BR^{d-k}$ for some $k\in\{1,...,d\}$
and impose Neumann boundary conditions.
The goal of this paper is to show that at low energy the particles will stick to the boundary of the domain.
In fact, the particles want to be close to as many boundary planes as possible.
In particular, they stick to the corner where all boundary planes intersect. 
Neumann boundary conditions can be interpreted as representing perfect mirrors. It is remarkable that while such boundary conditions are not sufficiently attractive to capture single particles, mutually bound pairs are always attracted to the boundary. 

In order to justify the picture of particles sticking to the boundary, we show that introducing a boundary plane lowers the ground state energy.
Then it is energetically favorable for the particles to localize at a finite distance to the new boundary plane.
Moving the particles away from that boundary plane would reduce the boundary effects and 
raise the energy to reach the previous ground state energy, which is strictly higher.
Since moving just one of the particles to infinity would increase the potential energy between them, both particles stick to the boundary.

This problem was already studied (for particles with equal masses) in the case $d=k=1$.
Kerner and M\"uhlenbruch \cite{kerner_two-particle_2017} considered a hard-wall interaction between the particles. 
(For a higher-dimensional version of this problem, which is different from the one we consider here, however, see \cite{bakharev_existence_2021}.)
More general interactions were studied  by Egger, Kerner and Pankrashkin in \cite{egger_bound_2020}.
Additionally, they showed that the Hamiltonian has only finitely many eigenvalues below the essential spectrum.
We show here that this also holds true for particles with different masses and all dimensions $d$ and numbers of boundary planes $k$. The finiteness of the number of bound states is a consequence of the fact that the effective attractive interaction with the boundary decays exponentially with distance, a decay that is inherited from the corresponding one of the ground state wave function in free space.

Let $x^a$ and $x^b$ be the coordinates of the particles.
The Hamiltonian of the system is
\begin{equation}
H=-\frac{1}{2m_a}\Delta_{x^a}-\frac{1}{2m_b}\Delta_{x^b}+V( x^a-x^b)
\end{equation}
acting in $L^2\left((0,\infty)^k \times \BR^{d-k}\right)\otimes L^2\left((0,\infty)^k \times \BR^{d-k}\right)$, where $V:\BR^d\to \BR$ is the interaction potential.
We change to relative and center-of-mass coordinates $y={x^a-x^b}$ and  $z=\frac{m_a x^a+m_b x^b}{M}$, where $M=m_a+m_b$ is the total mass.
The conditions $x^a_j >0$ and $x^b_j>0$ for $1\leq j\leq k$ result in the coordinates $(z_1,...,z_k,y_1,...,y_k)$ lying in the domain 
\begin{equation}
Q_k=\left\{(z_1,...,z_k,y_1,...,y_k) \in \BR^{2k}\ \vert\ \forall j \in \{1,...,k\}:  z_j>0\ \mathrm{and}\ -\frac{M}{m_b}z_j< y_j<\frac{M}{m_a}z_j\right\},
\end{equation}
while $(z_{k+1},...,z_d)$ and $(y_{k+1},...,y_d)$ lie in $\BR^{d-k}$.
In these coordinates, the Hamiltonian becomes
$H=-\frac{1}{2\mu} \Delta_{y}-\frac{1}{2M}\Delta_{z}+V( y),$
where $\mu= \frac{m_a m_b}{M}$ is the reduced mass.
Separating the variables $(z_{k+1},...,z_d)$ from the rest,
we write the Hamiltonian as $H=H_k\otimes \BI + \BI\otimes q$,
where $q=-\frac{1}{2M}\Delta$ on $H^2(\BR^{d-k})$ and
\begin{equation}
H_k=-\frac{1}{2\mu}\Delta_{y}-\frac{1}{2M}\sum_{j=1}^k \frac{\partial^2}{\partial z_j^2}+V( y)
\end{equation}
acting in $L^2(Q_k\times \BR^{d-k})$.
To be precise, we define the Hamiltonian $H_k$ via the quadratic form
\begin{equation}\label{def:hk}
h_k [\psi]=\int_{Q_k\times \BR^{d-k}}\left( \frac{1}{2\mu}\vert \nabla_{y} \psi\vert^2+\frac{1}{2M}\sum_{j=1}^k \left \vert \frac{\partial \psi}{\partial z_j}\right \vert^2+V( y) \vert \psi \vert^2\right) \dd z_1...\dd z_k \dd y_1...\dd y_d
\end{equation}
with domain $D[h_k]=H^1(Q_k\times \BR^{d-k})$.
Due to the free part of the kinetic energy $q$, the Hamiltonian $H$ has no discrete spectrum if $k<d$.
We remove this  free part and work with $H_k$ instead of $H$.

We impose the following conditions on the interaction potential $V$.
\begin{thm8}\label{assump_V_general}
We assume that
\begin{enumerate}[(i)]
\item $V=v+w$ for some $v \in L^{r}(\BR^d)$ and $w\in L^\infty (\BR^d)$, where 
\begin{align}
r=1\quad &\mathrm{if}\ d=1,\\
r>1\quad &\mathrm{if}\ d=2,\\
r\geq \frac{d}{2}\quad &\mathrm{if}\ d\geq 3,
\end{align} \label{V_reg}
\item the operator
$H_0=-\frac{1}{2\mu}\Delta_y+V(y)$ in $L^2(\BR^d)$ 
has a ground state $\psi_0$ with energy $E^0<0$, \label{H0GS} 
\item $\lim\inf_{\vert y\vert\to \infty}V(y)\geq 0$, \label{Vto0} 
\item  $V$ is invariant under permutation of the $d$ coordinates $(y_1, ..., y_d)\in \BR^d$. \label{V_permutation_invariant} 
\end{enumerate}
\end{thm8}

\begin{thm5}\label{why_assump}
Condition~(\ref{V_reg}) implies that in the quadratic form $h_k$ the interaction term is infinitesimally form bounded with respect to the kinetic energy, see Proposition~\ref{H_self-adj} in the Appendix.
The KLMN theorem (see e.g.~Theorem 6.24 in \cite{teschl_mathematical_2014}) then guarantees that there is a unique self-adjoint operator $H_k$ corresponding to $h_k$, which is bounded from below.
Assumption~(\ref{H0GS}) means that the particles form a bound state in free space.
Condition~(\ref{Vto0}) is a rather strong form of decay of the negative part at infinity. 
Presumably some weaker assumptions would be sufficient, but in our proofs this version is convenient. Also the assumptions on the positive part of $V$ can probably be relaxed.
Assumption~(\ref{V_permutation_invariant}) is imposed for convenience as it implies that it is irrelevant which coordinates are restricted, and without loss of generality we pick the first $k$. 
However, our methods easily extend to the general case.
\end{thm5}

Our first result is that the ground state energy strictly decreases upon adding a Neumann boundary  that cuts space in half, i.e.~when going from $k\to k+1$.
Moreover, the essential spectrum after dividing space starts at the previous ground state energy. 
\begin{thm3}\label{general_thm}
Let $V$ satisfy Assumptions~\ref{assump_V_general}.
Then for every $k\in \{1,...,d\}$, the bottom of the spectrum of the operator $H_k$ is an isolated eigenvalue $E^k=\inf \sigma(H_k)$.
Moreover, the essential spectrum of $H_k$ is $\sigma_{\textrm{\rm ess}}(H_k)=[E^{k-1},\infty)$.
In particular, the ground state energies form a decreasing sequence $E^d<E^{d-1}<...<E^0<0$.
\end{thm3}

Our second result is that the operators $H_k$ have only finitely many bound states.

\begin{thm3}\label{Hk_fin_ev}
Let $1\leq k\leq d$. 
Then $H_k$ has a finite number of eigenvalues below the essential spectrum.
\end{thm3}

In the one-dimensional case $d=k=1$ with equal masses $m_a=m_b$, Theorems~\ref{general_thm} and \ref{Hk_fin_ev} were  proved in \cite{egger_bound_2020}.
While we follow their main ideas, several new ingredients are needed to extend the results to general $d$ and $k$. In particular, 
the localization procedure in the proofs is more complicated and requires several additional steps.

\begin{thm5}
At various places it will be convenient to switch back to the particle coordinates in the first $k$ components, while keeping the relative coordinate in the last $d-k$ components.
We shall from now on use the notation $x^a=(x^a_1,...,x^a_k),\ x^b=(x^b_1,...,x^b_k)$ for the first $k$ components of the particle coordinates and $\tilde y=(y_{k+1},...,y_d)$ for the remaining components of the relative coordinate.
In this notation, $y=(x^a-x^b,\tilde y)$ and
\begin{equation}\label{hk_normal}
h_k [\psi]=\int_{[0,\infty)^{2k} \times \BR^{d-k}}\! \left( \frac{1}{2m_a}\vert \nabla_{x^a} \psi\vert^2+\frac{1}{2m_b}\vert \nabla_{x^b} \psi\vert^2+\frac{1}{2\mu}\vert \nabla_{\tilde y} \psi\vert^2+V( x^a-x^b, \tilde y) \vert \psi \vert^2\right)\dd x^a  \dd x^b \dd \tilde y
\end{equation}
with domain $D[h_k]=H^1((0,\infty)^{2k} \times \BR^{d-k})$.
\end{thm5}

\begin{thm5}\label{psi0pos}
By Corollary 5.1~in \cite{faris_quadratic_1972}, if $H_k$ has a ground state, it is non-degenerate and we can choose the corresponding wave function to be positive almost everywhere. 
\end{thm5}

The remainder of this paper is structured as follows.
Section~\ref{sec:gt} contains the proof of Theorem~\ref{general_thm}.
In Section~\ref{sec:fds}, we prove Theorem~\ref{Hk_fin_ev}.
The Appendix contains an explicit example for $d=1$ in \ref{sec:ex}, the proof of Lemma~\ref{h_l^N} in \ref{sec:pf_lea_hlN}, as well as technical details of the proofs in \ref{sec:tech-det}.
The exponential decay of Schr\"odinger eigenfunctions needed in the proof is discussed in Appendix \ref{sec:app_b} by Rupert L. Frank.

\section{Proof of Theorem~\ref{general_thm}} \label{sec:gt}

We shall prove the following two statements.
\begin{thm6}\label{esspec_general}
Let $k\in \{1,...,d\}$.
If $H_{k-1}$ has a ground state with energy $E^{k-1}\leq ... \leq E^0$ the essential spectrum of $H_k$ is $[E^{k-1},\infty)$.
\end{thm6}

\begin{thm6}\label{energy_seq}
Let $k\in \{1,...,d\}$.
If $H_{k-1}$ has a ground state $\psi_{k-1}$ with energy $E^{k-1}$
the spectrum of $H_k$ satisfies
\begin{equation}\label{eq:p22}
E^k = \inf \sigma(H_k)\leq  E^{k-1}- \frac{J^2 M}{8\mu^2}  \left(1+ 2\max\left\{\frac{m_a}{m_b},\frac{m_b}{m_a}\right\}\right)^{-1}<E^{k-1},
\end{equation}
where $J=\int_{Q_{k-1}\times \BR^{d-k+1}}\delta(y_k)\vert \psi_{k-1}\vert ^2 \dd z \dd y>0$ with $\delta$ the Dirac delta-function.
\end{thm6}
The assumption $E^{k-1}\leq ... \leq E^0$ in the first Proposition holds as a consequence of the second Proposition.
These two propositions combined yield Theorem~\ref{general_thm}.
\begin{proof}[Proof of Theorem~\ref{general_thm}]
We proceed by induction.
The claim is that $H_k$ has a ground state, and that the ground state energies form a strictly decreasing sequence $E^d<...<E^0$.
For $k=0$ the former is true by Assumption~\ref{assump_V_general}(\ref{H0GS}).
For the induction step we apply Propositions~\ref{esspec_general} and \ref{energy_seq}.
Assuming that the claim is true for $k-1$, Proposition~\ref{energy_seq} implies that $H_k$ has spectrum below $E^{k-1}$.
By Proposition~\ref{esspec_general} this part of the spectrum must consist of eigenvalues.
Since $H_{k}$ is bounded from below by Proposition~\ref{H_self-adj}, it must have a ground state. 
The ground state energy $E^k$ is strictly smaller than $E^{k-1}$ by Proposition~\ref{energy_seq}.
\end{proof} 

\subsection{Proof of Proposition~\ref{esspec_general}}

In order to compute the essential spectrum of $H_k$, we follow the proof of Proposition 2.1 in \cite{egger_bound_2020}.
For the inclusion $[E^{k-1},\infty)\subset \sigma_{\text{ess}}( H_k) $ we use Weyl's criterion (see Section 6.4 in \cite{teschl_mathematical_2014}). 
For the opposite inclusion, we bound the essential spectrum of $H_k$ from below by introducing additional Neumann boundaries.
They split the particle domain into several regions.
One of them is bounded, so it does not contribute to the essential spectrum. In another,  the interaction potential is larger than $E^{k-1}$, and hence there is no essential spectrum below $E^{k-1}$.
In the remaining regions, the Hamiltonian can be bounded from below by approximately $H_{k-1}\otimes \BI$.
For this operator the essential spectrum starts at $E^{k-1}$.

\begin{proof}[Proof of Proposition~\ref{esspec_general}]
For the inclusion $[E^{k-1},\infty)\subset \sigma_{\text{ess}}( H_k) $ we construct a Weyl sequence.
Remark~\ref{psi0pos} allows us to choose the ground state wave function $\psi_{k-1}$ of  $H_{k-1}$ to be normalized and positive almost everywhere. 
Let $l\in [0,\infty)$ and let $\tau:\BR \to \BR$ be a smooth function satisfying $0\leq \tau \leq 1$ with $\tau(x)=0$ for $x\leq 1$ and $\tau(x)=1$ for $x\geq 2$ .
Let us write $\delta=M/\max\{m_a,m_b\}$.
For integers $n\geq 5 $, choose $\varphi_n(z_1,...,z_k,y_1,...,y_d)=f_n(z_1,...,z_{k-1},y_1,...,y_d)g_n(z_k)$ for $(z,y)\in Q_{k}\times \BR^{d-k}$ with
\begin{equation}
f_n(z_1,...,z_{k-1},y_1,...,y_d)=\psi_{k-1}(z_1,...,z_{k-1},y_1,...,y_d)\tau (n-\vert y_k\vert/\delta)
\end{equation}
and
\begin{equation}
g_n(z_k)=\cos (l z_k) \tau (z_k-n)\tau (2n-z_k).
\end{equation}
Using the properties of $\tau$, we observe that $g_n(z_k)=\cos(l z_k)$ for $z_k\in[n+2,2n-2]$.
Moreover, for $\vert y_k\vert < \delta(n-2)$ we have $f_n=\psi_{k-1}$.
Note that for $(z,y)\in Q_k\times\BR^{d-k}$ with $z_k\geq n+2$, the variable $y_k$ can take all values satisfying $\vert y_k \vert \leq \delta(n+2)$.
Therefore,
\begin{equation}\label{normphin}
\lVert\varphi_n\rVert_{L^2(Q_k\times \BR^{d-k} )}^2\geq\left(\int_{Q_{k-1}\times [\delta(-n+2),\delta(n-2)]\times\BR^{d-k}} \psi_{k-1}^2\right)\left( \int_{n+2}^{2n-2} \cos^2(l z_k) \dd z_k\right).
\end{equation}
Since $\psi_{k-1}$ is normalized, the first integral converges to $1$ as $n\to \infty$.
The second integral is greater than some constant times $n$.
Thus, $\lVert\varphi_n\rVert_{L^2(Q_k\times \BR^{d-k} )}^2\geq C_1 n$ for some constant $C_1>0$.

Using the eigenvalue equation for $\psi_{k-1}$, we have 
\begin{equation}
\left(H_k- E^{k-1}-\frac{l^2}{2M} \right)\varphi_n= f_n \Psi_n + \Phi_n g_n 
\end{equation}
with
\begin{align}
\Psi_n(z_k)&=\frac{1}{M}l \sin(l z_k) \left[\tau'(z_k-n)\tau (2n-z_k) -\tau(z_k-n)\tau' (2n-z_k)\right] \\
& {}- \frac{1}{2M}\cos(l z_k) \left[\tau''(z_k-n)\tau (2n-z_k) -2\tau'(z_k-n)\tau' (2n-z_k)+\tau(z_k-n)\tau'' (2n-z_k)\right] \nonumber
\end{align}
and
\begin{equation}
\Phi_n(z_1,...,z_{k-1},y_1,...,y_d)=
\frac{1}{\delta\mu } \partial_{y_k}\psi_{k-1}\sgn (y_k)\tau'(n-\vert y_k \vert/\delta) -\frac{1}{2\delta^2\mu}\psi_{k-1} \tau''(n-\vert y_k \vert/\delta) .
\end{equation}
By choice of the function $\tau$, we have $\supp \Psi_n \subset [n+1,n+2]\cup [2n-2,2n-1]$ and $\supp \Phi_n\subset Q_{k-1}\times [\delta(-n+1),\delta(-n+2)]\cup [\delta(n-2),\delta(n-1)]\times \BR^{d-k}$.  
Since both $\tau'$ and $\tau''$ are bounded, there is a constant $C_2>0$ independent of $n$ such that $\vert\Phi_n\vert\leq C_2 \left(\vert \partial_{y_k}\psi_{k-1} \vert + \vert \psi_{k-1} \vert\right)$ and $\lVert \Psi_n\rVert_\infty\leq C_2$.
With the aid of the Schwarz inequality, we therefore have
\begin{align}\nonumber 
& \left\lVert\left(H_k-E^{k-1}-\frac{l^2}{2M}\right)\varphi_n\right\rVert_{L^2(Q_k\times \BR^{d-k} )}^2\\ \nonumber
& \leq 2  \int_{Q_{k-1} \times \BR^{d-k+1}} f_n^2 \int_{[n+1,n+2]\cup [2n-2,2n-1]} \Psi_n^2  
+ 2 \int_{Q_{k-1}\times  \BR^{d-k+1}} \Phi_n^2  \int_{n+1}^{2n-1} g_n^2 
 \\ 
& \leq 4 C_2^2  \left(1 + (n-2) \int_{Q_{k-1}\times [\delta(-n+1),\delta(-n+2)]\cup [\delta(n-2),\delta(n-1)]\times \BR^{d-k}} \left( (\partial_{y_k}\psi_{k-1})^2 +  \psi_{k-1}^2 \right) \right) 
\end{align}
where we used  $\lVert \psi_{k-1} \rVert_{L^2}=1$ in the last step. 
Since $\psi_{k-1} \in H^1(Q_{k-1}\times \BR^{d-k+1})$ we obtain
\begin{multline}
\lim_{n\to \infty}\frac{\lVert(H_k-E^{k-1}-\frac{l^2}{2M})\varphi_n\rVert_{L^2(Q_k\times \BR^{d-k} )}^2}{\lVert\varphi_n\rVert_{L^2(Q_k\times \BR^{d-k} )}^2}\\
\leq  \frac{4 C_2^2}{C_1}   \lim_{n\to \infty}  \int_{Q_{k-1}\times [\delta(-n+1),\delta(-n+2)]\cup [\delta(n-2),\delta(n-1)]\times \BR^{d-k}} \left( (\partial_{y_k}\psi_{k-1} )^2 +  \psi_{k-1}^2 \right) 
=0.
\end{multline}
By Weyl's criterion, we obtain $E^{k-1}+\frac{l^2}{2M}\in \sigma(H_k)$ for all $l\geq 0$.
Since the interval $[E^{k-1},\infty)$ has no isolated points, it belongs to the essential spectrum of $H_k$.

For the opposite inclusion $ \sigma_{\text{ess}}( H_k) \subset [E^{k-1},\infty)$, we partition the domain $Q_k \times \BR^{d-k}$ into $k+2$ subsets.
By Assumption~\ref{assump_V_general}(\ref{Vto0}) there is a number $L_0$ such that for all $y\in \BR^d$ with $\vert y \vert >L_0$ the potential satisfies $ V(y)> E^0$. 
For  $L>L_0$ and $1\leq l\leq k$ let
\begin{align}
\Omega_l&:=\left\{(z,y) \in Q_k \times \BR^{d-k}\ \Big\vert\  z_l >\frac{L}{\delta}, \vert y_l \vert <L, \forall 1\leq j<l: z_j<\frac{L}{\delta} \right\},\\
\Omega_{k+1}&:=\left\{(z,y) \in Q_k \times \BR^{d-k}\ \Big\vert\ \forall 1\leq j\leq k: z_j  < \frac{L}{\delta}, \forall j>k: \vert y_j \vert <L \right\} ,\\
\Omega_{k+2}&:=\Omega_0 \setminus \bigcup_{l=1}^{k+1}\overline{ \Omega_l} .
\end{align}
These sets are sketched in Figure~\ref{essspec_domains}.
The set $\Omega_{k+1}$ is bounded.
For $(z,y)\in \Omega_{k+2}$, we always have $\vert y \vert > L$.
Moreover, in $\Omega_l$ the range of $y_l$ is independent of $z_l$.
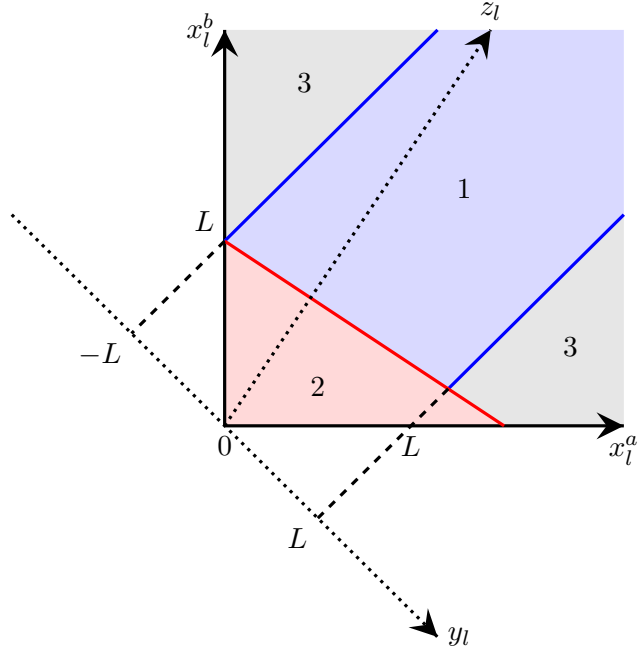
\begin{figure}
\centering

\begin{tikzpicture}[scale=0.35]
\coordinate (A) at  (3*14/5,3*14/5-7) {};

\fill[fill=red!15]    (0,0) --  (0,7)--(21/2,0);
\fill[fill=blue!15]     (A) -- (0,7) -- (8,15) -- (15,15)--(15,8);
\fill[fill=gray!20]    (0,7) -- ++ (0,8) -- ++(8,0);
\fill[fill=gray!20]     (21/2,0) -- (A)  -- (15,8) -- (15,0);


\draw (0,0) node[below]{$0$};
\draw (0,7) node[above left]{$L$};
\draw (0,15) node[left]{$x_l^b$};
\draw (7,0) node[below]{$ L$};
\draw (15,0) node[below]{$x_l^a$};
\draw (10,15) node[above]{$z_l$};
\draw (8,-8) node[right]{$y_l$};
\draw (-3.5,3.5) node[below left]{$-L$};
\draw (3.5,-3.5) node[below left]{$L$};

\draw (9,9) node{1};
\draw (3.5,1.5) node{2};
\draw (3,13) node{3};
\draw (13,3) node{3};

 \draw[black, very thick, decoration={markings, mark=at position 1 with {\arrow[scale=2,>=stealth]{>}}},
        postaction={decorate}]
 (0,0)--(15,0);
 \draw[black, very thick, decoration={markings, mark=at position 1 with {\arrow[scale=2,>=stealth]{>}}},
        postaction={decorate}]
 (0,0)--(0,15);

 \draw[blue, very thick]
 (0,7)--(8,15)
 (A) --(15,8);
 
  \draw[red, very thick]
 (0,7)--(21/2,0);

  \draw[black, very thick, dashed]
 (0,7)--(-3.5,3.5)
 (A)--(3.5,-3.5);

 \draw[black, very thick, dotted, decoration={markings, mark=at position 1 with {\arrow[scale=2,>=stealth]{>}}},
        postaction={decorate}]
 (0,0)--(10,15);
 \draw[black, very thick, dotted, decoration={markings, mark=at position 1 with {\arrow[scale=2,>=stealth]{>}}},
        postaction={decorate}]
 (-8,8)--(8,-8);
\end{tikzpicture} 
\caption{ In the case $d=k=1$, the areas labeled 1, 2, and 3 are precisely $\Omega_1,\Omega_2,\Omega_3$, respectively. In higher dimensions, region 1 (blue) is the domain of the $l$th component of $z$ and $y$ for $(z,y)\in \Omega_l$, $l\leq k$. In particular, the domain of $y_l$ is independent of $z_l$. The (red) triangular area 2 corresponds to the domain of $z_j$ and $y_j$ for $(z,y)\in \Omega_{l}$ and $j< l \leq k+1$.}
\label{essspec_domains}
\end{figure}

For $1\leq l \leq k+2$ we define the quadratic forms $a_l: H^1(\Omega_l) \to \BR$ as
\begin{equation}\label{al_prop1}
a_l[\psi]:=\int_{\Omega_l}\left(\frac{1}{2M}\vert \nabla_{z} \psi \vert^2+\frac{1}{2\mu}\vert \nabla_{y} \psi \vert^2+V( y)\vert \psi \vert^2\right)\dd z \dd y.
\end{equation}
For $1\leq l \leq k+1$, the potential term in $a_l$ is infinitesimally bounded with respect to the kinetic energy term, as will be shown in Lemma~\ref{a_self-adj_prop_1}.
For $a_{k+2}$ the potential is bounded from below.
Thus, by the KLMN theorem there is a corresponding self-adjoint operator $A_l$ for all $1\leq l \leq k+2$.
Let $A=\bigoplus_{l=1}^{k+2} A_l$.
There is an isometry $\iota: H^1(\Omega_0)\to \bigoplus_{l} H^1(\Omega_l)$, $\varphi \mapsto \{  \left. \varphi \right|_{\Omega_l}\}$.
Let $\{\varphi_n\}$ be a normalized Weyl sequence such that $\lim_{n\to \infty} \lVert (H_k-\inf\sigma_{\text{ess}}(H_k))\varphi_n\rVert=0$.
Then $\{ \iota(\varphi_n)\}$ is an orthonormal sequence with $\lim_{n\to \infty} \langle \iota(\varphi_n)\vert A\iota(\varphi_n)\rangle=\inf\sigma_{\text{ess}}(H_k)$.
By the min-max principle, 
\begin{equation}\label{eq:infsessHk}
\inf \sigma_{\text{ess}}(H_k)\geq \inf \sigma_{\text{ess}}\left(A\right)=\min_{l} \inf \sigma_{\text{ess}} (A_l). 
\end{equation}

We shall now analyze $\inf \sigma_{\text{ess}}(A_l)$ for all $1\leq l\leq k+2$.
Since $\Omega_{k+1}$ is a bounded Lipschitz domain, $H^1(\Omega_{k+1})$ is compactly embedded in $L^2(\Omega_{k+1})$ by the Rellich--Kondrachov theorem \cite{adams_sobolev_2003}.
Therefore, $A_{k+1}$ has compact resolvent and the spectrum of $A_{k+1}$ is discrete.
In $\Omega_{k+2}$, always at least one of the $y_j$ is larger than $L$.
Therefore, $\inf \sigma (A_{k+2})\geq \inf_{\vert y\vert >L} V(  y )\geq E^0$.

Consider now $A_l$ with $l\leq k$.
In order to separate the variable $z_l$  from the rest, let $q$ be the quadratic form $q[\varphi] =\frac{1}{2M}\int_{L/\delta}^\infty  \left \vert\varphi^\prime (z_l) \right \vert^2 \dd z_l$ with domain $H^1((L/\delta,\infty))$.
The remaining variables lie in
\begin{multline}
\Omega_{k-1}^{L,l}:= \left\{(z_1,...,\widehat z_l,...,z_k,y_1,...,y_d) \in \BR^{d+k-1}\ \Big\vert\  \forall 1\leq j<l: 0<z_j<\frac{L}{\delta}, \forall j>l: z_j>0, \right. \\
\left. \forall 1\leq j \neq l \leq k: -\frac{M}{m_b}z_j< y_j < \frac{M}{m_a} z_j,  \vert y_l \vert <L  \right\}
\end{multline}
where the hat means that the $z_l$ variable is omitted.
Note that for $L\to \infty$ the set $\Omega_{k-1}^{L,l}$ becomes $Q_{k-1}\times \BR^{d-k+1}$ with $l$ and $k$ components swapped.
Define the quadratic form 
\begin{equation}\label{h_l^N_def}
h^{L,l}_{k-1}[\psi]=\int_{\Omega_{k-1}^{L,l}}\left(\frac{1}{2M}\sum_{ \genfrac {}{} {0pt} {2} {j=1}{j\neq l}}^{k}\left \vert \frac{\partial \psi}{\partial z_j}\right \vert^2+\frac{1}{2\mu}  \left \vert\nabla_{y}\psi\right \vert^2 +V( y)\vert \psi \vert^2\right)\dd z_1 ... \widehat{\dd z_l} ... \dd z_k \dd y
\end{equation}
with domain $D[h_{k-1}^{L,l}]=H^1(\Omega_{k-1}^{L,l})$.
In Lemma~\ref{a_self-adj_prop_1} we show that there is a self-adjoint operator $H_{k-1}^{L,l}$ corresponding to the quadratic form $h_{k-1}^{L,l}$.
By Assumption~\ref{assump_V_general}(\ref{V_permutation_invariant}), the quadratic form $h_{k-1}^{L,l}$ resembles $h_{k-1}$ with $l$ and $k$ components swapped, up to the constraints imposed by the finite number $L$.

We can decompose
\begin{equation}\label{al=hl+q}
a_l=h^{L,l}_{k-1} \otimes \id + \id \otimes q.
\end{equation}
It is well-known that the self-adjoint operator corresponding to $q$ has purely essential spectrum $[0,\infty)$.
Therefore, we obtain $\inf \sigma_{\text{ess}} (A_l)=\inf \sigma(H_{k-1}^{L,l})$.
Using localization arguments, one can easily prove the following.
\begin{thm2}\label{h_l^N}
Let $1\leq l\leq k \leq d$ and assume that $E^{k-1}\leq...\leq E^0$. The self-adjoint operator $H^{L,l}_{k-1}$ defined through the quadratic form (\ref{h_l^N_def}) satisfies $\liminf_{L\to \infty}\inf \sigma (H^{L,l}_{k-1})\geq E^{k-1}$.
\end{thm2}

The proof of Lemma~\ref{h_l^N} is rather straightforward and follows similar arguments as 
in the one-dimensional case in Proposition A.5~in \cite{egger_bound_2020}.
For completeness, we carry it out in Appendix~\ref{sec:pf_lea_hlN}.

Collecting all estimates and applying (\ref{eq:infsessHk}), we see that $\inf \sigma_{\text{ess}}(H_k)\geq \min\{E^0, \inf \sigma (H^{L,l}_{k-1})\}$ for all $L>L_0$.
With Lemma~\ref{h_l^N} and since $E^0\geq E^{k-1}$, it follows that $ \sigma_{\text{ess}}( H_k) \subset [E^{k-1},\infty)$.
\end{proof}

\subsection{Proof of Proposition~\ref{energy_seq}}
The goal is to find a trial function $\psi$ such that $(\psi, H_k \psi)<E^{k-1}\lVert \psi \rVert_2^2$.
Then $\inf \sigma(H_k)<E^{k-1}$ by the min-max principle.

We denote the ground state of $H_{k-1}$ by $\psi_{k-1}$ and choose it normalized and positive a.e. (see Remark~\ref{psi0pos}).
Since we expect the ground state of $H_k$ to stick to the boundary, we pick the trial function
\begin{equation}
\psi(z_1,...,z_k,y_1,...,y_d)=\psi_{k-1}(z_1,...,z_{k-1},y_1,...,y_d)e^{-\gamma z_k}
\end{equation}
for $\gamma>0$.
We start with a preliminary computation.
\begin{thm2}\label{seq_norm}
Let $f(y_k)=\chi_{(-\infty,0)}(y_k)e^{-2\gamma m_b \vert y_k \vert /M}+\chi_{(0,\infty)}(y_k)e^{-2\gamma m_a  \vert y_k \vert /M}$, where $\chi$ denotes the characteristic function.
We have
\begin{equation}
A:= \frac{1}{2} (f \psi_{k-1},\psi_{k-1})=\gamma \lVert \psi \rVert^2_2 .
\end{equation}
\end{thm2}
\begin{proof}
Carrying out the integration over $z_k$, we have
\begin{align}
\lVert \psi \rVert^2_2 &=  \int_{Q_{k-1}\times  \BR^{d-k+1}} \dd z_1...\dd z_{k-1} \dd y \int_{0}^\infty \dd z_k \ \chi_{\{-\frac{M}{m_b}z_k<y_k<\frac{M}{m_a}z_k\}}\psi_{k-1}^2(z_1,...,z_{k-1},y_1,...,y_d) e^{-2\gamma z_k} \nonumber \\
&=\frac{1}{2\gamma} \int_{Q_{k-1} \times \BR^{d-k+1}} \dd z_1...\dd z_{k-1} \dd y \  \psi_{k-1}^2(z_1,...,z_{k-1},y_1,...,y_d) f(y_k) \nonumber \\
&= \frac{1}{2\gamma} (f\psi_{k-1},\psi_{k-1})=\frac{1}{\gamma} A.
\end{align}
\end{proof}

\begin{proof}[Proof of Proposition~\ref{energy_seq}]
We have
\begin{multline}\label{seq_exp}
h_k[\psi]= \int_{ Q_k \times \BR^{d-k}}\dd z_1... \dd z_k \dd y_1...\dd y_d \left(\frac{1}{2M}\vert \nabla_z \psi_{k-1} \vert^2+\frac{1}{2\mu}\vert \nabla_y \psi_{k-1}\vert^2 \right. \\
\left. +\frac{\gamma^2}{2M}  \psi_{k-1}^2 +V( y)  \psi_{k-1}^2 \right)e^{-2\gamma z_k} .
\end{multline}
We rewrite this as
\begin{multline}\label{seq_exp2}
h_k [\psi]=\frac{\gamma^2 \lVert \psi \rVert^2_2}{2M}+\int_{ Q_{k-1} \times \BR^{d-k+1}} \dd z_1...\dd z_{k-1} \dd y_1...\dd y_d \\
\int_0^\infty \dd z_k\chi_{\{-\frac{M}{m_b}z_k<y_k<\frac{M}{m_a}z_k\}} \left(\frac{1}{2M}\vert \nabla_z \psi_{k-1} \vert^2
+\frac{1}{2\mu} \vert \nabla_y \psi_{k-1}\vert^2 +V( y)  \psi_{k-1}^2 \right)e^{-2\gamma z_k} .
\end{multline}
Integrating over $z_k$ as in the proof of Lemma~\ref{seq_norm}, we obtain
\begin{multline}
h_k [\psi]=\frac{\gamma^2\lVert \psi \rVert^2_2}{2M}+\frac{1}{2\gamma } \int_{ Q_{k-1} \times \BR^{d-k+1}} \dd z_1...\dd z_{k-1} \dd y_1...\dd y_d  \left(\frac{1}{2M}\vert \nabla_z \psi_{k-1} \vert^2\right.\\
\left.+\frac{1}{2\mu}\vert \nabla_y \psi_{k-1}\vert^2 +V( y)  \psi_{k-1}^2 \right)f(y_k) .
\end{multline}
We pull the function $f$ into the gradients and write
\begin{multline}
h_k [\psi]=\frac{\gamma^2 \lVert \psi \rVert^2_2}{2M}+ \frac{1}{2\gamma } \int_{ Q_{k-1}\times \BR^{d-k+1}} \left(\frac{1}{2M} {\nabla_z (f \psi_{k-1})} \nabla_z \psi_{k-1} +\frac{1}{2\mu}  {\nabla_y ( f \psi_{k-1})}\nabla_y \psi_{k-1} \right.\\
\left.  +\frac{\gamma}{\mu M} \left(-m_b \chi_{(-\infty,0)} e^{-2 \gamma \frac{m_b}{M} \vert y_k \vert}+m_a \chi_{(0,\infty)} e^{-2\gamma \frac{m_a}{M} \vert y_k \vert } \right) \psi_{k-1}\partial_{y_k} \psi_{k-1}+V( y)  f \psi_{k-1}^2 \right) .
\end{multline}
Let us write $h_k[\cdot,\cdot]$ for the sesquilinear form associated to the quadratic form $h_k$.
The previous equation reads
\begin{equation}
h_k [\psi]=\frac{\gamma^2 \lVert \psi \rVert^2_2}{2M}+ \frac{1}{2\gamma } h_{k-1} [f \psi_{k-1},\psi_{k-1}] + B,
\end{equation}
where
\begin{equation}
B= \frac{1}{2 \mu M } \int_{ Q_{k-1}\times \BR^{d-k+1}}\left(-m_b \chi_{(-\infty,0)} e^{-2 \gamma \frac{m_b}{M} \vert y_k \vert}+m_a \chi_{(0,\infty)} e^{-2\gamma \frac{m_a}{M} \vert y_k \vert } \right) { \psi_{k-1}} \partial_{y_k} \psi_{k-1}.
\end{equation}
Since $\psi_{k-1}$ is the minimizer of the functional $\frac{h_{k-1}[\phi]}{\lVert \phi \rVert^2_2}$, for all functions $g\in H^1(Q_{k-1}\times \BR^{d-k+1})$ it holds that $h_{k-1}[g,\psi_{k-1}]=E^{k-1}(g,\psi_{k-1}).$
With $g= f \psi_{k-1}$ and Lemma~\ref{seq_norm}, we obtain
\begin{equation}\label{ek_comp_1}
h_k [\psi]=\left(\frac{\gamma^2}{2M}+E^{k-1}\right)\lVert \psi \rVert^2_2+B.
\end{equation}
We now simplify the integral in $B$.
By the Sobolev embedding theorem (Theorem 4.12 in \cite{adams_sobolev_2003}), the restriction of an $H^1$-function to a hyperplane is an $L^2$-function.
Therefore, one can restrict the function $\psi_{k-1}$ to $y_k=0$ and obtain a finite number $J:= \int_{Q_{k-1}\times \BR^{d-k}}\left(\left. \psi_{k-1}\right\vert_{y_k=0}\right)^2$.
Integration by parts with respect to $y_k$ gives
\begin{align}
2 \mu M  B=& - m_b \int_{ Q_{k-1} \times (-\infty,0)\times \BR^{d-k}} e^{-2\gamma \frac{m_b}{M} \vert y_k\vert}  { \psi_{k-1}}\partial_{y_k} \psi_{k-1} \nonumber\\
&+ m_a \int_{ Q_{k-1} \times (0,\infty)\times \BR^{d-k}}  e^{-2\gamma \frac{m_a}{M}\vert y_k\vert}  { \psi_{k-1}}\partial_{y_k} \psi_{k-1} \nonumber  \\ 
=&- \frac{m_b}{2}\int _{Q_{k-1} \times \BR^{d-k}}\left(\left. \psi_{k-1}\right\vert_{y_k=0}\right)^2+ \gamma \frac{m_b^2}{M} \int_{ Q_{k-1}\times (-\infty,0)\times \BR^{d-k}} e^{-2\gamma \frac{m_b}{M} \vert y_k\vert } \psi_{k-1}^2 \nonumber\\
&-\frac{m_a}{2}\int _{Q_{k-1} \times \BR^{d-k}}\left(\left. \psi_{k-1}\right\vert_{y_k=0}\right)^2+\gamma \frac{m_a^2}{M} \int_{ Q_{k-1}\times (0,\infty)\times \BR^{d-k}}  e^{-2\gamma \frac{m_a}{M} \vert y_k\vert }\psi_{k-1}^2 \nonumber \\
=&-\frac{M}{2}J \nonumber\\
&+\frac{\gamma}{M} \int_{ Q_{k-1}\times \BR^{d-k+1}} \left( m_b^2 \chi_{(-\infty,0)}(y_k) e^{-2\gamma \frac{m_b}{M} \vert y_k\vert }+ m_a^2\chi_{(0,\infty)}(y_k)  e^{-2\gamma \frac{m_a}{M} \vert y_k\vert }\right) \psi_{k-1}^2.
\end{align}
The last integral is bounded from above by $2 \max\{m_a^2,m_b^2\} A$. 
With (\ref{ek_comp_1}), Lemma~\ref{seq_norm} and the min-max principle we obtain
\begin{equation}
\inf \sigma(H_k)\leq \frac{h_k[\psi]}{\lVert \psi \rVert^2_2}\leq E^{k-1}+\frac{\gamma }{A} \left( \left(\frac{1}{2}+ \max\left\{ \frac{m_a}{m_b},\frac{m_b}{m_a} \right\} \right) \frac{\gamma A}{M} -\frac{J}{4\mu}\right).
\end{equation}
This holds for all $\gamma>0$. Minimizing with respect to $\gamma$ yields
\begin{equation}
\inf \sigma(H_k)\leq E^{k-1}-\frac{J^2 M}{32 \mu^2 A^2} \left(1+ 2  \max\left\{ \frac{m_a}{m_b},\frac{m_b}{m_a} \right\} \right)^{-1}.
\end{equation}
Moreover, since $\psi_{k-1}$ is normalized we have
\begin{equation}
A= \frac{1}{2} \int_{Q_{k-1}\times \BR^{d-k+1}} f \psi_{k-1}^2 \leq \frac{1}{2} \int_{Q_{k-1}\times \BR^{d-k+1}}   \psi_{k-1}^2 =\frac{1}{2}. 
\end{equation}
This yields \eqref{eq:p22}. 

We are left with showing that $J>0$.
Suppose that $J=0$.
Define a new function $\widetilde \psi_{k-1}=\psi_{k-1}\left(\chi_{y_k<0}-\chi_{y_k>0}\right)$.
Since $J=0$, the function $\widetilde \psi_{k-1} \in H^1(Q_{k-1}\times \BR^{d-k+1})$. 
Moreover, $\widetilde \psi_{k-1}$ is a ground state of $H_{k-1}$ because $\frac{h_{k-1}[\widetilde \psi_{k-1}]}{\lVert \widetilde \psi_{k-1} \rVert^2_2}= \frac{h_{k-1}[\psi_{k-1}]}{\lVert \psi_{k-1} \rVert^2_2}$.
Since $\psi_{k-1}$ and $\widetilde \psi_{k-1}$ are linearly independent, this contradicts the uniqueness of the ground state (Remark~\ref{psi0pos}). 
Hence, $J>0$ and $\inf \sigma(H_k)<E^{k-1}$.
\end{proof}

\section{Finiteness of the Discrete Spectrum}\label{sec:fds}

In this section we shall give the proof of Theorem~\ref{Hk_fin_ev}. An important ingredient will be the exponential decay of the ground state wave function $\psi_{k}$ of $H_k$. In fact, the Agmon estimate (Corollary 4.2. in \cite{agmon_lectures_1983}) implies that for any $a<\sqrt{\inf \sigma_{\text{ess}}(H_k)-E^k}$ we have 
\begin{equation}\label{eq:agmon1}
\int_{Q_k\times \BR^{d-k}} \vert \psi_k \vert^2 e^{2a\sqrt{2M\vert z \vert^2+2\mu \vert y \vert^2}} \dd z \dd y <\infty.
\end{equation}
Strictly speaking, the assumptions on the interaction potential stated in \cite{agmon_lectures_1983} are slightly stronger than ours. 
However, the Agmon estimate only requires $V$ to be form-bounded with respect to the kinetic energy with form bound less than $1$, as shown in Theorem~\ref{expdecay} in Appendix~\ref{sec:app_b} by Rupert Frank.
As we argue in Proposition~\ref{H_self-adj}, this is the case given Assumptions~\ref{assump_V_general}.

In order to derive \eqref{eq:agmon1} from Theorem~\ref{expdecay}, we remove the boundaries in the particle domain via mirroring and consider the operator $\widetilde H_k$ acting on $H^1(\BR^{d+k})$ (see Proposition~\ref{mirror_details}).
It suffices to prove the exponential decay for the ground state $\widetilde \psi_k$ of $\widetilde H_k$.
We rescale the variables to remove the masses in front of the Laplacians using the unitary transform $U\varphi (z,y)= \sqrt{2M}^k \sqrt{2\mu}^d \varphi(\sqrt{2M}z, \sqrt{2\mu} y)$ on $H^1(\BR^{d+k})$ .
Switching to relative and center of mass coordinates and writing $\widetilde V(z,y)=V((\vert x^a_j\vert -\vert x^b_j \vert)_{j=1}^k, \tilde y)$ and $\widetilde V_U(z,y)=\widetilde V(z/\sqrt{2M},y/\sqrt{2\mu})$ we have
\begin{equation}
\widetilde H_k=-\frac{1}{2M}\Delta_z -\frac{1}{2\mu} \Delta_y+\widetilde V =U \left( -\Delta_z - \Delta_y+\widetilde V_U \right)U^\dagger.
\end{equation}
The ground state $\varphi_k$ of $-\Delta_z - \Delta_y+\widetilde V_U$ satisfies $\widetilde \psi_k= U \varphi_k$.
For any $a<\sqrt{\inf \sigma_{\text{ess}}(H_k)-E^k}=\sqrt{\inf \sigma_{\text{ess}}(\widetilde H_k)-E^k}$ we thus have
\begin{equation}
\int_{\BR^{d+k}} \vert \widetilde \psi_k \vert^2 e^{2a\sqrt{2M\vert z \vert^2+2\mu \vert y \vert^2}} \dd z \dd y =\int_{\BR^{d+k}} \vert \varphi_k\vert^2 e^{2a\sqrt{\vert z \vert^2+ \vert y \vert^2}} \dd z \dd y<\infty
\end{equation}
by Theorem~\ref{expdecay}.
 Hence \eqref{eq:agmon1} holds. 

\begin{thm1}
Let $n\in \BZ^{\geq 0}$ and $A$ be a self-adjoint operator with corresponding quadratic form $a$.
We define
\begin{equation}
E_{n}(A):=\inf_{ \genfrac {}{} {0pt} {2} {V \subset D[a]}{ \dim V=n+1} }\sup_{ \genfrac{}{}{0pt}2{\varphi\in V}{\varphi \neq 0}} \frac{a[\varphi]}{ \\ \lVert \varphi \rVert^2}.
\end{equation}
By the min-max principle, if $n$ is larger than the number of eigenvalues below the essential spectrum, we have $E_n(A)=\inf \sigma_{\text{ess}}(A)$.
Otherwise, $E_{n-1}$ is the $n$-th eigenvalue of $A$ below the essential spectrum counted with multiplicities.
\end{thm1}

\begin{thm1}
For a self-adjoint operator $A$ and a number $\lambda\in\BR$, let $N(A,\lambda)$ denote the number of eigenvalues in $(-\infty,\lambda)$ if $ \sigma_{\text{ess}}(A)\cap (-\infty, \lambda)= \emptyset$.
Otherwise, set $N(A,\lambda)=\infty$.
When $N(A,\lambda)\neq 0$, one can write
\begin{equation}
N(A,\lambda)=\sup \left\{{n\in \BZ^{\geq 1}} \vert  E_{n-1}(A)<\lambda \right\} .
\end{equation}
\end{thm1}

In the case $k=d=1$, Theorem~\ref{Hk_fin_ev} was already shown in~\cite{egger_bound_2020}.
We generalize the proof using similar ideas.
The overall strategy is to construct localized operators $A$ and bound $N(H_k,E^{k-1})$ using $N(A,E^{k-1})$.
The localized operators fall into three categories. 
First, they can have compact resolvent or second, the corresponding potential is larger than $E^{k-1}$.
In these cases, the number of eigenvalues below $E^{k-1}$ is certainly finite (or even zero).
In the third category, the operator is of the form $\BI\otimes H_{k-1}-\frac{1}{2M}\Delta_{z_j}\otimes \BI -K$, where $K$ is a well behaved error term.
One estimates this operator by projecting onto $L^2(\BR)\otimes \psi_{k-1}$ and its orthogonal complement.
This reduces the problem to a one-dimensional operator.
Then, \eqref{eq:agmon1}  and the Bargmann estimate~\cite{berezin_schrodinger_1991} imply that the number of eigenvalues is finite.

\begin{proof}[Proof of Theorem~\ref{Hk_fin_ev}]
Let $\chi_1,\chi_2:\BR\to [0,1]$ and $\chi_3:\BR^2\to [0,1]$ be continuously differentiable functions satisfying $\chi_1(t)=0$ for $t\geq 2$, $\chi_1(t)=1$ for $t\leq 1$, $\chi_1(t)^2+\chi_2(t)^2=1$ for all $t$ and $\chi_3(s,t)^2+\chi_2(s)^2\chi_2(t)^2=1$ for all $t$ and $s$. Note that for $j=1,2,3$ we have $\lVert( \nabla \chi_j)^2 \rVert_\infty< \infty $.

Let $\Omega_0=(0,\infty)^{2k}\times \BR^{d-k}$.
The boundary of the particle domain consists of $k$ orthogonal $d-1$-dimensional hyperplanes.
We start by localizing into two separate regions, distinguishing whether there is a particle close to all the hyperplanes, or whether both particles are far from some hyperplane.
For $R>0$, let
\begin{align}
\Omega_1&=\left\{(x^a,x^b, \tilde y)\in \Omega_0\vert x^a \in (0,2R)^k\ \mathrm{or}\   x^b \in (0,2R)^k \right\}\nonumber \\
&=\left\{(x^a,x^b, \tilde y)\in \Omega_0 \vert \max\{x_1^a,...,x_k^a\}< 2R\ \mathrm{or}\  \max\{x_1^b,...,x_k^b\}< 2R \right\},\\
\Omega_2&=\left\{(x^a,x^b, \tilde y)\in\Omega_0 \vert x^a \not\in [0,R]^k\ \mathrm{and}\   x^b \not\in [0,R]^k \right\} \nonumber \\
&=\left\{(x^a,x^b, \tilde y)\in\Omega_0 \vert \max\{x_1^a,...,x_k^a\}> R\ \mathrm{and}\  \max\{x_1^b,...,x_k^b\}> R \right\}.
\end{align}
We define the functions
\begin{align}
f_1^R(x^a,x^b)&=\chi_3\left(\frac{\max\{x_1^a,...,x_k^a\}}{R},\frac{\max\{x_1^b,...,x_k^b\}}{R}\right),\\
f_2^R(x^a,x^b)&=\chi_2\left(\frac{\max\{x_1^a,...,x_k^a\}}{R}\right)\chi_2\left(\frac{\max\{x_1^b,...,x_k^b\}}{R}\right).
\end{align}
Note that for all functions $\varphi \in L^2(\Omega_0)$ we have support $\supp f_j^R \varphi \subset \Omega_j$ .
By the IMS localization formula we have for all $\varphi \in H^1(\Omega_0)$ that
\begin{equation}
h_k[f_1^R \varphi]+h_k[f_2^R \varphi]= h_k[\varphi]+\int_{(0,\infty)^{2k} \times \BR^{d-k} }W_R \vert\varphi \vert^2 \, \dd x^a \dd x^b \dd \tilde y \,,
\end{equation}
where
\begin{align}\nonumber
 W_R(x^a,x^b,\tilde y) & = \frac{1}{R^2}\left[\frac{1}{2m_a}(\nabla_{x^a} \chi_3) \left(\frac{\max\{x_1^a,...,x_k^a\}}{R},\frac{\max\{x_1^b,...,x_k^b\}}{R}\right)^2 \right.\\ \nonumber
& \qquad\qquad +\frac{1}{2m_b}(\nabla_{x^b} \chi_3) \left(\frac{\max\{x_1^a,...,x_k^a\}}{R},\frac{\max\{x_1^b,...,x_k^b\}}{R}\right)^2\\ \nonumber
&\qquad\qquad +\frac{1}{2m_a}\chi'_2\left(\frac{\max\{x_1^a,...,x_k^a\}}{R}\right)^2\chi_2\left(\frac{\max\{x_1^b,...,x_k^b\}}{R}\right)^2\\ 
& \qquad\qquad + \left.\frac{1}{2m_b}\chi_2\left(\frac{\max\{x_1^a,...,x_k^a\}}{R}\right)^2\chi'_2\left(\frac{\max\{x_1^b,...,x_k^b\}}{R}\right)^2  \right]. \label{def:WR}
\end{align}
Note that there is a constant $c_1>0$ such that $\lVert W_R \rVert_\infty \leq \frac{c_1}{R^2}$.
For $j=1,2$, define the quadratic forms
\begin{align}\nonumber
a_j[\varphi]
& = \int_{\Omega_j} \biggl(  \frac{1}{2m_a}\vert \nabla_{x^a} \varphi\vert^2+\frac{1}{2m_b} \vert \nabla_{x^b} \varphi\vert^2+\frac{1}{2\mu}\vert \nabla_{\tilde y} \varphi\vert^2 \\
& \qquad\qquad +\left(V(x^a-x^b, \tilde y)-W_R(x^a,x^b, \tilde y)\right) \vert \varphi \vert^2\biggl) \dd x^a \dd x^b \dd \tilde y \label{aj_nr_ev_k}
\end{align}
with domains
\begin{align}
D[a_1]&=\left\{\varphi \in H^1(\Omega_0) \vert \varphi(x^a,x^b,\tilde y)=0 \quad \mathrm{if}\quad  \max\{x_1^a,...,x_k^a\}\geq2R\ \mathrm{and}\ \max\{x_1^b,...,x_k^b\}\geq2R  \right\}, \\
D[a_2]&=\left\{\varphi \in H^1(\Omega_0) \vert  \varphi(x^a,x^b,\tilde y)=0 \quad \mathrm{if}\quad  \max\{x_1^a,...,x_k^a\}\leq R\ \mathrm{or}\ \max\{x_1^b,...,x_k^b\}\leq R  \right\}.
\end{align}
For all quadratic forms $a_j$ in this proof, let $A_j$ denote the corresponding self-adjoint operator.
In Lemma~\ref{a_self-adj_gen}, we verify that these operators exist.
For $\varphi\in D[h_k]$, the restriction of the function $f_j^R \varphi$ to $\Omega_j$ belongs to $D[a_j]$.
With $(f_1^R)^2+(f_2^R)^2=1$, it follows that $h_k[\varphi]=a_1[f_1^R \varphi] + a_2[f_2^R \varphi]$.
Let $\hat A$ denote the operator $\hat A = A_1 \oplus A_2$.
The map $J: H^1(\Omega_0)\to H^1(\Omega_0)\oplus H^1(\Omega_0), \varphi \mapsto (f_1^R \varphi, f_2^R \varphi)$ is an $L^2$-isometry and thus injective.
By the min-max principle, we have
\begin{multline}
E_n(H_k)=\inf_{ \genfrac{}{}{0pt}2{V \subset D[h_k] }{\dim V=n+1} }\sup_{\genfrac{}{}{0pt}2{\varphi\in V}{\varphi \neq 0}} \frac{h_k[\varphi]}{  \lVert \varphi \rVert^2_{L^2(\Omega_0)}}=\inf_{ \genfrac{}{}{0pt}2 {V \subset D[h_k]}{\dim V=n+1} }\sup_{\genfrac{}{}{0pt}{2}{\varphi\in V}{\varphi \neq 0}} \frac{\hat a[J\varphi]}{ \lVert J\varphi \rVert^2_{L^2(\Omega_0)\oplus L^2(\Omega_0)}}\\
=\inf_{\genfrac{}{}{0pt}2{V \subset JD[h_k]}{\dim V=n+1}}\sup_{\genfrac{}{}{0pt}{2}{\varphi\in V}{\varphi \neq 0}} \frac{\hat a[\varphi]}{ \lVert \varphi \rVert^2_{L^2(\Omega_0)\oplus L^2(\Omega_0)}}\geq \inf_{\genfrac{}{}{0pt}2{V \subset D[\hat a]}{\dim V=n+1}}\sup_{\genfrac{}{}{0pt}2{\varphi\in V}{\varphi \neq 0}} \frac{\hat a[\varphi]}{ \lVert \varphi \rVert^2_{L^2(\Omega_0)\oplus L^2(\Omega_0)}}=E_n(\hat A)
\end{multline}
 for all $n \in \BZ^{\geq 0}$.
Thus, $N(H_k, E^{k-1})\leq N(\hat A, E^{k-1})=N( A_1, E^{k-1})+N( A_2, E^{k-1})$.

Let
\begin{align}
\tilde \Omega_{1,\text{int}}&=\left\{(x^a,x^b,\tilde y)\in \Omega_0 \vert  (x^a-x^b,\tilde y) \in (-R,R)^d \right\} \quad \mathrm{ and}\\
\tilde \Omega_{1,\text{ext}}&=\left\{(x^a,x^b, \tilde y)\in \Omega_0 \vert   (x^a-x^b, \tilde y)\not\in [-R,R]^d  \right\}.
\end{align}
Moreover, let $\Omega_{1,\bullet}=\tilde \Omega_{1,\bullet} \cap \Omega_1$ for $\bullet\in\{\text{int},\text{ext}\}$.
Define quadratic forms $a_{1,\text{int}},a_{1,\text{ext}}$ through expression~(\ref{aj_nr_ev_k}) with domain
\begin{equation}\label{a_1_bullet}
D[a_{1,\bullet}]=\left\{\varphi \in H^1(\tilde \Omega_{1,\bullet}) \vert \varphi(x^a,x^b,\tilde y)=0 \ \mathrm{if}\  \max\{x_1^a,...,x_k^a\}\geq2R\ \mathrm{and}\ \max\{ x_1^b,...,x_k^b\}\geq2R  \right\},
\end{equation}
for $\bullet\in\{\text{int},\text{ext}\}$.
Again, there is an isometry $D[a_1]\to D[a_{1,\text{int}}]\oplus  D[a_{1,\text{ext}}], \varphi \mapsto ( \varphi\vert_{\tilde \Omega_{1,\text{int}}} ,  \varphi\vert_{\tilde \Omega_{1,\text{ext}}})$, and therefore, $N(A_1,E^{k-1})\leq 
N(A_{1,\text{int}},E^{k-1})+N(A_{1,\text{ext}},E^{k-1})$.
Since the negative part of $V$ vanishes at infinity by Assumption~\ref{assump_V_general}(\ref{Vto0}) and since $\lVert W_R \rVert_\infty \leq \frac{c_1}{R^2}$, there is a $R_0>0$ such that for $R\geq R_0$ and $\vert (x^a-x^b,\tilde y)\vert\geq R_0$ we have $V(x^a-x^b,\tilde y)-W_R(x^a,x^b,\tilde y)> E^{k-1}$.
Choosing $R\geq R_0$, we have $N(A_{1,\text{ext}},E^{k-1})=0$.
Since $\Omega_{1,\text{int}}$ is a bounded Lipschitz domain, $A_{1,\text{int}}$ has purely discrete spectrum.
As $A_{1,\text{int}}$ is bounded from below, we have $N(A_{1,\text{int}}, E^{k-1})< \infty$.

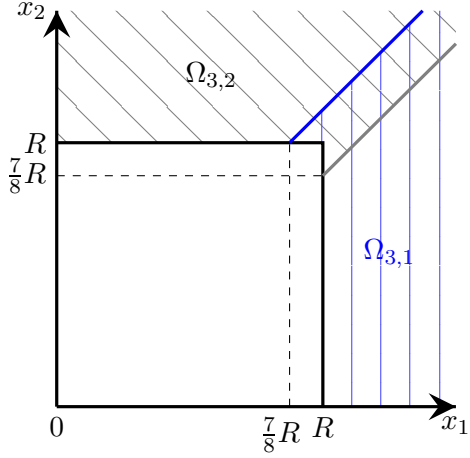
\begin{figure}
\centering

\begin{tikzpicture}[scale=0.35]

\pattern[pattern color=gray!100, pattern=my north west lines]    (10,10) -- ++ (-10,0) -- ++(0,5) -- ++(15,0)--++(0,-1.25)--++(-5,-5);
\pattern[pattern color=blue!70, pattern=blue up lines]     (10,0) -- ++ (0,10) --++ (-1.25,0) --++ (5,5) --++(1.25, 0) -- ++(0,-15);


\draw (0,0) node[below]{$0$};
\draw (0,10) node[left]{$R$};
\draw (0,15) node[left]{$x_2$};
\draw (10,0) node[below]{$R$};
\draw (15,0) node[below]{$x_1$};

\draw (5.8,12.5) node[color=black]{$\Omega_{3,2}$};
\draw (12.5,5.8) node[color=blue]{$\Omega_{3,1}$};

 \draw[black, very thick, decoration={markings, mark=at position 1 with {\arrow[scale=2,>=stealth]{>}}},
        postaction={decorate}]
 (0,0)--(15,0);
 \draw[black, very thick, decoration={markings, mark=at position 1 with {\arrow[scale=2,>=stealth]{>}}},
        postaction={decorate}]
 (0,0)--(0,15);

 \draw[black, very thick] (0,10)--(10,10)--(10,0);

 \draw[blue, very thick] (8.75,10)--(13.75,15);
 \draw[gray, very thick] (10,8.75)--(15,13.75);

 \draw[dashed] (0,8.75)--(10,8.75);
\draw[dashed] (8.75,0)--(8.75,10);
\draw (8.4,0) node[below]{$\frac{7}{8}R$};
\draw (0,8.75) node[left]{$\frac{7}{8}R$};

\end{tikzpicture} 
\caption{Let $k=2$. In $\Omega_2$ both $x^a$ and $x^b$ lie outside the square $(0,R)^2$. 
 If $x^a$ lies below the upper diagonal, the configuration belongs to $\Omega_{3,1}$. 
If $x^a$ lies above the lower diagonal, the configuration belongs to $\Omega_{3,2}$.}
\label{nr_ev_k_2}
\end{figure}

We are left with showing that $N(A_2,E^{k-1})<\infty$.
For $k=1$, wave functions in the support of $A_2$ are localized away from the boundary. Effectively, the boundary has thus disappeared and one can directly make a comparison with $H_{k-1}= H_0$. 
For $k>1$, the domain $\Omega_2$ is more complicated and we need to continue localizing in order to effectively eliminate one of the boundary planes. 
For now, assume $k>1$ and let $r=R/8$.
We localize $x^a$ in the $k$ sectors
\begin{equation} \label{Omega_3,j}
\Omega_{3,j}=\{(x^a,x^b,\tilde y)\in \Omega_2 \vert x^a_j > \max\{x^a_1,...,x^a_k\} - r\} \quad \mathrm{for}\ 1\leq j\leq k.
\end{equation}
In the sector $\Omega_{3,j}$, the largest component of $x^a$ is $x^a_j$ up to the constant $r$.
The domains are sketched in Figure~\ref{nr_ev_k_2} for the case $k=2$.
For the localization, we need functions $f_{3,j}^r$ on $\Omega_2$ which are supported in $\Omega_{3,j}$, satisfy $\sum_{j=1}^k (f^r_{3,j})^2=1$, and their derivatives scale as $1/r$.
We construct auxiliary functions $f_{3,j}$ corresponding to the case $r=1$ and set 
\begin{equation}\label{fr_3j}
f^r_{3,j}(x^a,x^b,\tilde y)=f_{3,j}({x^a}/{r}).
\end{equation}
The idea behind the construction of the auxiliary functions is as follows.
We want that $f_{3,1}$ equals $1$ on $\Omega_{3,1}$ apart from the boundary region which overlaps with other $\Omega_{3,j}$.
The expression $\max\{x^a_{2},...,x^a_k\}-x_1^a$ measures the distance to the boundary of $\Omega_{3,1}$ and is large outside $\Omega_{3,1}$.
Hence, to define $f_{3,1}$, we apply $\chi_1$ to this expression (up to some constants).
For the sum condition to hold, the remaining $f_{3,j}$ will contain the corresponding factor $\chi_2$.
This $\chi_2$ factor takes care of the behavior at the boundary towards large $x_1^a$.
For the next function $f_{3,2}$, we proceed analogously to before, but ignoring the $x_1^a$ direction.
Inductively, for $x^a \in (0,\infty)^k$ and $1\leq j \leq k-1$ we define
\begin{align}
f_{3,j}(x^a)&=\chi_1\left(\frac{k}{2}\left({\max\{x^a_{j+1},...,x^a_k\}-x_j^a}\right)+\frac{3}{2}\right) \prod_{l=1}^{j-1} \chi_2\left(\frac{k}{2}\left({\max\{x^a_{l+1},...,x^a_k\}-x_l^a}\right)+\frac{3}{2}\right), \nonumber\\
f_{3,k}(x^a)&=\prod_{l=1}^{k-1} \chi_2\left(\frac{k}{2}\left({\max\{x^a_{l+1},...,x^a_k\}-x_l^a}\right)+\frac{3}{2}\right), \label{f_3j}
\end{align}
where the product in the first line has to be understood as $1$ for $j=1$. 
Note that for all $1\leq j\leq k$ the derivatives are bounded, i.e.~$\lVert(\nabla f_{3,j})^2\rVert_\infty < \infty$.
By construction, we have $\sum_{j=1}^k (f_{3,j})^2=1$.
That the functions $f^r_{3,j}$ indeed have the correct support is the content of the following Lemma, which is proved at the end of this section.

\begin{thm2}\label{supp_fj}
For $1\leq j \leq k$, the functions $f^r_{3,j}$ defined through (\ref{fr_3j}) and (\ref{f_3j}) satisfy
\begin{equation}
\supp f^r_{3,j} \cap \Omega_2\subset \overline{ \Omega_{3,j}}.
\end{equation}
Moreover,
\begin{equation}
\supp \nabla f^r_{3,j} \cap \Omega_2 \subset \{(x^a,x^b,\tilde y)\in \Omega_2 \vert \max\{x_1^a,...,\widehat{ x^a_j},...,x^a_k\}-r \leq x^a_j \leq \max\{x_1^a,...,\widehat{ x^a_j},...,x^a_k\}+r\},
\end{equation}
where $\widehat{x^a_j}$ means that this variable is omitted.
\end{thm2}

By the IMS formula, we have for all $\varphi \in D[a_2]$
\begin{equation}
\sum_{j=1}^k a_2[f_{3,j}^r \varphi]=a_2[ \varphi]+\int_{\Omega_2} F_r(x^a,x^b,\tilde y) \vert \varphi \vert^2 \dd x^a \dd x^b\dd \tilde y,
\end{equation}
where
\begin{equation}\label{def:Fr}
F_r(x^a,x^b,\tilde y)=\frac{1}{r^2}\sum_{j=1}^k \frac{1}{2m_a}(\nabla f_{3,j})^2\left(x^a/r\right).
\end{equation}
For $1\leq j\leq k$, define the quadratic forms
\begin{multline}\label{a_3_j}
a_{3,j}[\varphi]= \int_{\Omega_{3,j}} \biggl(  \frac{1}{2m_a} \vert \nabla_{x^a} \varphi\vert^2+ \frac{1}{2m_b}\vert \nabla_{x^b} \varphi\vert^2+\frac{1}{2\mu}\vert \nabla_{\tilde y} \varphi\vert^2 \\
+ \left(V(x^a-x^b,\tilde y)-W_R(x^a,x^b,\tilde y)-F_r(x^a,x^b,\tilde y) \right) \vert \varphi \vert^2  \biggl) \dd x^a \dd x^b \dd \tilde y
\end{multline}
with domains
\begin{multline}
D[a_{3,j}]= \Big\{ \varphi \in H^1(\Omega_0) \vert\varphi(x^a,x^b,\tilde y)=0 \quad \mathrm{if}\quad  \max\{x_1^a,...,x_k^a\}\leq R\ \mathrm{or}\ \max\{x_1^b,...,x_k^b\} \leq R  \\
  \mathrm{or}\  x^a_j \leq \max\{x^a_1,...,x^a_k\} - r\Big\}.
\end{multline}
Again we have $N(A_2,E^{k-1})\leq \sum_{j=1}^k N(A_{3,j},E^{k-1})$.
We will show that $N(A_{3,k},E^{k-1})<\infty$.
For $1\leq j <k $, by Assumption~\ref{assump_V_general}(\ref{V_permutation_invariant}) the same argument with vector components $k\leftrightarrow j $ swapped gives $N(A_{3,j},E^{k-1})<\infty$.

We localize $x^b$ close and far from the domain of $x^a$.
Define the sets
\begin{align}
\Omega_4&=\left\{(x^a,x^b,\tilde y)\in \Omega_{3,k} \vert x_k^b> \max\{x^b_1,...,x^b_{k-1}\}-4r\right\} \quad \mathrm{and} \\
\Omega_5&=\left\{(x^a,x^b,\tilde y)\in \Omega_{3,k} \vert x_k^b< \max\{x^b_1,...,x^b_{k-1}\}-2r  \right\}. \label{Omega_5}
\end{align}
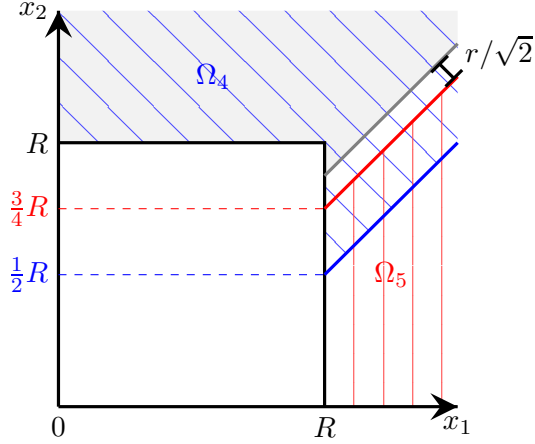
\begin{figure}
\centering
\begin{tikzpicture}[scale=0.35]

\fill[fill=gray!10]    (10,8.75) -- ++ (0,1.25) --++ (-10,0) --++ (0,5) --++(15, 0) -- ++(0,-1.25);
\pattern[pattern color=blue!70, pattern=blue north west lines]   (10,10) -- ++ (-10,0) -- ++(0,5) -- ++(15,0)--++(0,-5)--++(-5,-5);
\pattern[pattern color=red!70, pattern=red up lines]    (10,0) -- ++ (0,7.5) -- ++ (5,5) -- ++(0,-12.5) ;


\draw (0,0) node[below]{$0$};
\draw (0,10) node[left]{$R$};
\draw (0,15) node[left]{$x_2$};
\draw (10,0) node[below]{$R$};
\draw (15,0) node[below]{$x_1$};

\draw (0,5) node[left, color=blue]{$\frac{1}{2} R$};
\draw (0,7.5) node[left, color=red]{$\frac{3}{4} R$};

\draw (5.8,12.5) node[color=blue]{$\Omega_4$};
\draw (12.5,5) node[color=red]{$\Omega_5$};

 \draw[black, very thick, decoration={markings, mark=at position 1 with {\arrow[scale=2,>=stealth]{>}}},
        postaction={decorate}]
 (0,0)--(15,0);
 \draw[black, very thick, decoration={markings, mark=at position 1 with {\arrow[scale=2,>=stealth]{>}}},
        postaction={decorate}]
 (0,0)--(0,15);

 \draw[black, very thick] (0,10)--(10,10)--(10,0);

 \draw[gray, very thick] (10,8.75)--(15,13.75);
 \draw[red, very thick] (10,7.5)--(15,12.5);
 \draw[blue, very thick] (10,5)--(15,10);

 \draw[red, dashed] (0,7.5)--(10,7.5);
 \draw[blue, dashed] (0,5)--(10,5);

\draw[black, very thick, |-|](14.9,12.4)--++(135:0.88);
\draw (14.9,12.4) node[above right]{$r/\sqrt{2}$};

\end{tikzpicture} 
\caption{In $\Omega_{3,2}$, the first particle's coordinate $x^a$ lies in the shaded area, while  the second particle at $x^b$ lies outside the square $(0,R)^2$. If $x^b$ lies above the lowest  diagonal (blue), the configuration belongs to $\Omega_4$. If $x^b$ lies below the middle diagonal (red), the configuration belongs to $\Omega_5$. Note that for any configuration in $\Omega_5$, the particles are separated by at least distance $r/\sqrt{2}$. }
\label{nr_ev_k_3}
\end{figure}
For $k=2$, they are sketched in Figure~\ref{nr_ev_k_3}.
Let $f_4^r(x^b)=\chi_1\left(\frac{\max\{x^b_1,...,x^b_{k-1}\}-x^b_k}{2r}\right)$ and $f_5^r(x^b)=\chi_2\left(\frac{\max\{x^b_1,...,x^b_{k-1}\}-x^b_k}{2r}\right)$.
By the IMS formula, we have for all $\varphi \in D[a_{3,k}]$
\begin{equation}
a_{3,k}[f_4^r \varphi]+a_{3,k}[f_5^r \varphi]=a_{3,k}[ \varphi]+\int_{\Omega_{3,k}} G_r(x^a,x^b,\tilde y) \vert \varphi \vert^2 \dd x^a \dd x^b \dd \tilde y,
\end{equation}
where
\begin{equation}\label{def:Gr}
G_r(x^a,x^b, \tilde y)=\frac{1}{4r^2m_b}\left[ \chi'_1\left(\frac{\max\{x^b_1,...,x^b_{k-1}\}-x^b_k}{2r}\right)^2+ \chi'_2\left(\frac{\max\{x^b_1,...,x^b_{k-1}\}-x^b_k}{2r}\right)^2\right].
\end{equation}
For $j=4,5$, define the quadratic forms
\begin{multline}\label{aj_4_5}
a_j[\varphi]= \int_{\Omega_j} \biggl(  \frac{1}{2m_a} \vert \nabla_{x^a} \varphi\vert^2+\frac{1}{2m_b} \vert \nabla_{x^b} \varphi\vert^2+\frac{1}{2\mu} \vert \nabla_{\tilde y} \varphi\vert^2  \\
+\left(V(x^a-x^b,\tilde y)-W_R(x^a,x^b,\tilde y)-F_r(x^a,x^b,\tilde y)-G_r(x^a,x^b,\tilde y) \right) \vert \varphi \vert^2   \biggl) \dd x^a \dd x^b \dd\tilde y
\end{multline}
with domains
\begin{multline}
D[a_4]=\left\{\varphi \in H^1(\Omega_0) \vert  \varphi(x^a,x^b,\tilde y)=0 \quad \mathrm{if}\quad   \max\{x^a_1,...,x^a_k\}\leq R\ \mathrm{or}\  \max\{x^b_1,...,x^b_k\}\leq R \right. \\
 \left. \ \mathrm{or}\     x_k^a \leq \max\{x^a_1,...,x^a_{k-1}\}-r \ \mathrm{or}\   x_k^b\leq  \max\{x^b_1,...,x^b_{k-1}\}-4r  \right\},
\end{multline}
\begin{multline}
D[a_5]=\left\{\varphi \in H^1(\Omega_0) \vert\varphi(x^a,x^b,\tilde y)=0 \quad \mathrm{if}\quad  \max\{x^a_1,...,x^a_k\}\leq R\ \mathrm{or}\  \max\{x^b_1,...,x^b_k\}\leq R \right. \\
 \left. \ \mathrm{or}\     x_k^a\leq \max\{x^a_1,...,x^a_{k-1}\}-r \ \mathrm{or}\   x_k^b\geq  \max\{x^b_1,...,x^b_{k-1}\}-2r   \right\}.
\end{multline}
Again, we have $N(A_{3,k},E^{k-1})\leq N(A_4,E^{k-1})+N(A_5,E^{k-1})$.

For $(x^a,x^b,\tilde y)\in \Omega_5$, we claim that 
\begin{equation}\label{3.32}
\vert (x^a-x^b,\tilde y)\vert \geq r/\sqrt{2}=R/(8\sqrt{2}). 
\end{equation}
Let $l$ be the index such that $x^b_l=\max\{x^b_1,...,x^b_{k-1}\}$.
We estimate 
\begin{equation}\label{3.33}
\vert (x^a-x^b,\tilde y)\vert^2 \geq (x_l^a-x_l^b)^2+(x_k^a-x_k^b)^2 
\geq \frac{1}{{2}}\left(x_l^a-x_k^a -x_l^b+ x_k^b\right)^2 .
\end{equation}
Since $\max\{x^a_1,...,x^a_{k-1}\}\geq x^a_l$ we have in the set $\Omega_5$ (see (\ref{Omega_5}) and (\ref{Omega_3,j})) 
\begin{equation}\label{3.34}
x_k^a>x^a_l-r \quad \text{and} \quad x_k^b<x^b_l-2r \qquad  \Leftrightarrow \qquad x_l^a-x_k^a<r \quad \text{and} \quad x^b_l-x_k^b>2r.
\end{equation}
Combining this with (\ref{3.33}) yields (\ref{3.32}).
Moreover, we have $\lVert W_R \rVert_\infty+\lVert F_r \rVert_\infty+\lVert G_r \rVert_\infty \leq \frac{c_2}{R^2}$.
By Assumption~\ref{assump_V_general}(\ref{Vto0}), there is $R_1>0$ such that for $R>R_1$ we have $a_5 > E^{k-1}$.
Choosing $R$ large enough, we thus have $N(A_5,E^{k-1})=0$.

For $k=1$, we set $F_r=G_r=0$ and $a_4=a_2$.
For any choice of $k\geq 1$, we now just need to show $N(A_4,E^{k-1})<\infty$.
At the boundaries which constrain the $k$th component of $x^a$ and $x^b$, the operator $A_4$ has Dirichlet boundary conditions.
The idea is to extend the domain of $x^a_k$ and $x^b_k$ to $\BR$, which leads to the new operator $\hat A_4$ defined below.
In $\hat A_4$, the boundary hyperplane in the $k$th direction has disappeared.
This makes it possible to compare the operator $\hat A_4$ to the Hamiltonian $H_{k-1}$ of the problem with $k-1$ boundary hyperplanes.
Let us write $K_R=(W_R+F_r+G_r)\chi_{(0,\infty)^{2k}\times \BR^{d-k}}$. 
Let $\hat \Omega_4=\left( (0,\infty)^{k-1}\times \BR \right)^2\times \BR^{d-k}$ and define the quadratic form
\begin{multline}\label{a4_hat}
\hat{a}_4[\varphi]= \int_{\hat \Omega_4} \biggl( \frac{1}{2m_a}\vert \nabla_{x^a} \varphi\vert^2+ \frac{1}{2m_b}\vert \nabla_{x^b} \varphi\vert^2+\frac{1}{2\mu}  \vert \nabla_{\tilde y} \varphi\vert^2 \\
+ \left(V(x^a-x^b,\tilde y)-K_R(x^a,x^b,\tilde y)\right) \vert \varphi \vert^2  \biggl) \dd x^a \dd x^b \dd \tilde y
\end{multline}
with domain
$D[\hat a_4]= H^1(\hat \Omega_4) $.
We have $N(A_4, E^{k-1})\leq N(\hat A_4, E^{k-1})$.

Let us change to relative and center-of-mass coordinates $y=(x^a-x^b,\tilde y)$ and $z=\frac{m_a x^a+m_b x^b}{M}$.
Then
\begin{multline}
\hat{a}_4[\varphi]= \int_{\BR}\dd z_k \int_{Q_{k-1}\times \BR^{d-k+1}} \dd z_1..\dd z_{k-1} \dd y \biggl(  \frac{1}{2\mu}\vert  \nabla_{y} \varphi\vert^2+ \frac{1}{2M} \vert \nabla_{z} \varphi\vert^2  \\
+ \left[V(y)-K_R\left(z+\frac{m_b}{M}(y_1,...,y_k),z-\frac{m_a}{M}(y_1,...,y_k),\tilde y\right)\right] \vert \varphi \vert^2 \biggl)
\end{multline}
with $D[\hat{a}_4]=H^1(\BR \times Q_{k-1}\times \BR^{d-k+1})$.
Note that we can separate $z_k$ from the other variables and write the corresponding operator as $\hat A_4=\BI\otimes H_{k-1}-\frac{1}{2M}\Delta_{z_k}\otimes \BI -K_R$.
Recall that $H_{k-1}$ has the ground state $\psi_{k-1}$ with energy $E^{k-1}$.
Let $\Pi$ denote the orthogonal projection onto $L^2(\BR)\otimes \psi_{k-1}$ in $L^2(\BR\times Q_{k-1}\times \BR^{d-k+1})$, and $\Pi^\perp:= \BI-\Pi$.
For $\varphi\in H^1(\BR\times Q_{k-1}\times \BR^{d-k+1})$ both $\Pi\varphi$ and $\Pi^\perp \varphi$ belong to $H^1(\BR \times Q_{k-1}\times \BR^{d-k+1})$.
We have
\begin{equation}
\hat a_4[\varphi]=\hat a_4[\Pi \varphi]+\hat a_4[\Pi^\perp \varphi]-2K_R[\Pi^\perp \varphi,\Pi \varphi],
\end{equation}
where 
\begin{multline}
K_R[\varphi,\psi]=\int_{\BR\times Q_{k-1}\times \BR^{d-k+1}} \overline{\varphi (z,y)}K_R\left(z+\frac{m_b}{M}(y_1,...,y_k),z-\frac{m_a}{M}(y_1,...,y_k),\tilde y\right)\psi(z,y)\\
 \dd z_k \dd z_1... \dd z_{k-1}\dd y.
\end{multline}
Using the Schwarz inequality, we estimate
\begin{equation}
\vert 2K_R[\Pi^\perp \varphi,\Pi \varphi]\vert
\leq R \lVert K_R \Pi \varphi \rVert_{L^2(\BR\times Q_{k-1}\times \BR^{d-k+1})}^2 +\frac{1}{R} \lVert \Pi^\perp \varphi\rVert_{L^2(\BR\times  Q_{k-1}\times \BR^{d-k+1})}^2.
\end{equation}
Since $E^{k-1}$ is a discrete and non-degenerate eigenvalue of $H_{k-1}$, we have $E_1^{k-1}=\inf (\sigma(H_{k-1})\setminus \{E^{k-1}\})>E^{k-1}$, and 
 $(\BI \otimes h_{k-1}) [\Pi^\perp \varphi]\geq E_1^{k-1} \lVert \Pi^\perp \varphi\rVert_{L^2(\BR\times Q_{k-1} \times \BR^{d-k+1})}^2$.
Together with the positivity of $-\Delta_{z_k}\otimes \BI$  and  $\lVert K_R\rVert_\infty\leq \frac{c_2}{R^2}$  it follows that
\begin{equation}
\hat a_4[\Pi^\perp \varphi]\geq 
\left(E_1^{k-1}-\frac{c_2}{R^2}\right) \lVert \Pi^\perp \varphi\rVert_{L^2(\BR\times Q_{k-1}\times \BR^{d-k+1})}^2.
\end{equation}
In total, we have
\begin{equation}
\hat a_4[\varphi]
\geq \hat a_4[\Pi \varphi]-R \lVert K_R \Pi \varphi \rVert_{L^2(\BR\times Q_{k-1}\times \BR^{d-k+1})}^2
+\left(E_1^{k-1}-\frac{1}{R}-\frac{c_2}{R^2}\right)\lVert \Pi^\perp \varphi\rVert_{L^2(\BR\times Q_{k-1}\times \BR^{d-k+1})}^2.
\end{equation}
We choose $R$ large enough such that $E_1^{k-1}-E^{k-1}>\frac{1}{R}+\frac{c_2}{R^2}$. 
Let $B_1$ be the self-adjoint operator corresponding to 
\begin{equation}\label{b1}
b_1[\varphi]= \hat a_4[\varphi]-R \lVert K_R \varphi \rVert_{L^2(\BR\times Q_{k-1}\times \BR^{d-k+1})}^2 \end{equation} in $\ran \Pi$.
Then $N(\hat A_4,E^{k-1})\leq N(B_1, E^{k-1})$ by the min-max principle. 

We can write any function $\varphi \in \ran \Pi$ as $\varphi(z,y)=f(z_k)\psi_{k-1}(z_1,...,z_{k-1},y)$ for some $f\in H^1(\BR)$.
Integrating over $z_1,...,z_{k-1},y$, we have
\begin{equation}
\hat a_4[f\otimes \psi_{k-1}]=\int_{\BR} \left( \frac{1}{2M} \vert f'(z_k)\vert^2+(E^{k-1}-U_R(z_k))f(z_k)^2 \right)\dd z_k,
\end{equation}
where 
\begin{multline}
U_R(z_k)= \int_{Q_{k-1}\times \BR^{d-k+1}} K_R\left(z+\frac{m_b}{M}(y_1,...,y_k),z-\frac{m_a}{M}(y_1,...,y_k),\tilde y\right) \psi_{k-1}(z_1,...z_{k-1},y)^2 \\
\dd z_1...\dd z_{k-1} \dd y.
\end{multline}
Moreover,
\begin{equation}
\lVert K_R ( f \otimes \psi_{k-1} )\rVert_{L^2(\BR\times Q_{k-1}\times \BR^{d-k+1})}^2=\int_{\BR} V_R(z_k) f(z_k)^2 \dd z_k
\end{equation}
with 
\begin{multline}
V_R(z_k)= \int_{Q_{k-1}\times \BR^{d-k+1}} K_R\left(z+\frac{m_b}{M}(y_1,...,y_k),z-\frac{m_a}{M}(y_1,...,y_k),\tilde y\right)^2 \psi_{k-1}(z_1,...,z_{k-1},y)^2\\
 \dd z_1...\dd z_{k-1} \dd y.
\end{multline}
Let $Z_R=U_R+ R V_R$.
With 
\begin{equation}\label{b3}
b_2[f]=\int_{\BR}\left( \frac{1}{2M}\vert f'(z)\vert^2 -Z_R(z)f(z)^2\right) \dd z,
\end{equation} we can write $b_1[f\otimes \psi_{k-1}]=E^{k-1} \lVert f \rVert_{L^2(\BR)}^2 + b_2[f]$.
Therefore, $N(B_1, E^{k-1})=N(B_2,0)$.

In the following, we bound the function $Z_R$ from above by an exponentially decaying function.
With this bound it is easy to see that $N(B_2,0)<\infty$ using 
 e.g.~the Bargmann estimate (see Chapter 2, Theorem 5.3 in~\cite{berezin_schrodinger_1991}).
This concludes the proof of $N(H_k,E^{k-1})<\infty$.

To bound $Z_R$, first use that $K_R$ is bounded to obtain 
\begin{equation}\label{Z_bound}
Z_R(z_k)\leq \left(\lVert K \rVert_\infty + R\lVert K \rVert_\infty^2  \right) I(z_k),
\end{equation}
 where
\begin{equation}
I(z_k) = \int_{Q_{k-1}\times \BR^{d-k+1}}\chi_{\supp K_R} (z,y)  \psi_{k-1}^2  \dd z_1...\dd z_{k-1} \dd y.
\end{equation}
By construction, $I(z_k)=0$ for $z_k<0$. 
We shall show that $I(z_k)$ decays exponentially for $z_k\geq 0$.
In fact, if $z_k$ is large and $K_R(z,y)\neq 0$, then necessarily one of the remaining coordinates $z_1,...,z_{k-1},y_1,...,y_d$ has to be large as well. 
This is essentially the content of the following Lemma.
\begin{thm2}\label{alpha_estimate}
Let $a>0$. For $z_k \geq 2 R$ 
the function 
\begin{equation}
\alpha(z,y)=e^{a\sqrt{2M\vert z_1 \vert^2+..+2M\vert z_ {k-1}\vert^2+2\mu\vert y \vert^2}} \chi_{\supp K_R}(z,y) 
\end{equation}
satisfies $\alpha(z,y)\geq e^{a c(z_k-2R)} \chi_{\supp K_R}(z,y)$ with $c= \sqrt{2M} ( 1 + 2 \max\{ \frac{m_a}{m_b}, \frac{m_b}{m_a} \} )^{-1/2}$.
\end{thm2}

The Agmon estimate \eqref{eq:agmon1} tells us that there is a constant $a>0$ such that 
\begin{equation}
c_3 := \int_{Q_{k-1}\times \BR^{d-k+1}}  \psi_{k-1}^2 e^{a\sqrt{2M \vert z_1 \vert^2+...+2M\vert z_{k-1} \vert^2+2\mu \vert y \vert^2}} \dd z_1...\dd z_{k-1} \dd y<\infty.
\end{equation}
We apply Lemma~\ref{alpha_estimate} with this constant $a$ and conclude that  $ \chi_{\supp K_R}(z,y)\leq e^{-c_4(z_k-2R)}\alpha(z,y)$ for $z_k\geq 2R$ and suitable constant $c_4>0$. 
In particular, 
\begin{align}\nonumber 
I(z_k) & \leq  e^{-c_4 (z_k-2R)} \int_{Q_{k-1}\times \BR^{d-k+1}}\alpha(z,y) \psi_{k-1}(z_1,...,z_{k-1},y)^2 \dd z_1...\dd z_{k-1}\dd y\\
 & \leq  c_3 e^{-c_4 (z_k-2R)}  .
\end{align}
for $z_k \geq 2R$. 
Recall that $Z_R$ vanishes on $(-\infty,0)$ and $\lVert Z_R \rVert_\infty <\infty$.
With  (\ref{Z_bound})  we thus conclude the desired exponentially decaying bound.
\end{proof}

It remains to give the proof of Lemmas~\ref{supp_fj} and~\ref{alpha_estimate}.

\begin{proof}[Proof of Lemma~\ref{alpha_estimate}] Recall the definitions of $W_R$, $F_r$ and $G_r$ in \eqref{def:WR}, \eqref{def:Fr} and \eqref{def:Gr}, respectively. 
Since $\supp K_R \subset\ \supp W_R \cup\ \supp F_r \cup\ \supp G_r$, we estimate $\alpha$ on each of these three sets.  
In $\supp W_R$, at least one particle is close to the corner, i.e. in the hypercube $(0,2R)^k$.
If $z_k$ is large, this means that the two particles are far apart and $y_k$ is large.
To be  precise, 
using $x^a_j=z_j+\frac{m_b}{M}y_j$ and $x^b_j=z_j-\frac{m_a}{M}y_j$ we have
\begin{align}\nonumber
\supp W_R & \subset \left\{(z,y) \in Q_k\times \BR^{d-k} \vert 0\leq \frac{z_k+\frac{m_b}{M}y_k}{R} \leq 2 \ \mathrm{or}\ 0\leq \frac{z_k-\frac{m_a}{M}y_k}{R} \leq 2\right\}\\
 & \subset \left\{(z,y) \in Q_k\times \BR^{d-k} \vert z_k-2 R \leq \frac{\max\{m_a,m_b\}}{M}\vert y_k \vert \right\} .
\end{align}
For $(z,y) \in \supp W_R$ with $z_k\geq 2R$, we therefore have 
\begin{equation}
M \sum_{j=1}^{k-1} |z_j|^2 + \mu \sum_{j=1}^k |y_k|^2 \geq \frac{\mu M^2(z_k-2R)^2}{\max\{m_a^2,m_b^2\}}= \frac{ M (z_k - 2R)^2 }{\max\{ \frac{m_a}{m_b}, \frac{m_b}{m_a} \} },
\end{equation}
which implies the desired bound on $\alpha$.

For $k=1$, both $F_r$ and $G_r$ are identically zero, hence to estimate $\alpha$ on their support we can restrict our attention to the case $k>1$. Observe that in $\supp F_r$  every coordinate $x_j^a$ for $1\leq j \leq k$ is smaller than or similar in magnitude to the largest of the other coordinates $x_i^a$, $i\neq j$; in particular, this applies to $j=k$.  
Intuitively, for large $z_k$ either $x_k^a$ or $|y_k|$ needs to be large. 
If $x_k^a$ is large, also some other $x_j^a$ with $j<k$ has to be large.
Phrased precisely, by Lemma~\ref{supp_fj} we have
\begin{multline}
\supp F_r \subset \\
\bigcup_{j=1}^k\left\{(z,y) \in Q_k\times \BR^{d-k} \vert \max_{\substack{1\leq l\leq k,\\ l\neq j}}\{z_l+\frac{m_b}{M}y_l\}-r \leq z_j+\frac{m_b}{M}y_j  \leq \max_{\substack{1\leq l\leq k,\\ l\neq j}}\{z_l+\frac{m_b}{M}y_l\}+r \right\} \\
\subset \left\{(z,y) \in Q_k\times \BR^{d-k} \vert z_k -r \leq -\frac{m_b}{M}y_k+ \max_{1\leq j\leq k-1}\{\frac{m_b}{M}y_j+z_j\}\right\} =: S_F.
\end{multline}
The constraint in $S_F$ can be written as $z_k-r\leq (\sqrt{M} z, \sqrt{\mu}y)\cdot e$ for a vector $e\in \BR^{k+d}$.
A simple Schwarz inequality therefore shows that on the set $S_F$ we have
\begin{equation}
M \sum_{j=1}^{k-1} |z_j|^2 + \mu \sum_{j=1}^k |y_k|^2 \geq \frac{ (z_k - r)^2 }{ \lVert e \rVert^2 }= \frac{ M (z_k - r)^2 }{ 1+ 2 \frac{m_b}{m_a} }
\end{equation}
as long as $z_k \geq r$, which yields the desired bound on $\alpha$.  

Similarly to the previous case, in $\supp G_r$ the coordinate $x_k^b$ is of similar magnitude as the largest of the other coordinates $x_j^b$.
We have
\begin{align}\nonumber 
\supp G_r & \subset \left\{(z,y) \in Q_k\times \BR^{d-k} \vert 2r\leq{\max_{1\leq j\leq k-1}\{z_j-\frac{m_a}{M}y_j\}+\frac{m_a}{M}y_k-z_k}\leq 4r \right\}\\
& \subset \left\{(z,y) \in Q_k\times \BR^{d-k} \vert z_k +2r \leq \max_{1\leq j\leq k-1}\{z_j-\frac{m_a}{M}y_j\}+\frac{m_a}{M}y_k\right\}=:S_G.
\end{align}
Analogously to before, on the set $S_G$ we have 
\begin{equation}
M \sum_{j=1}^{k-1} |z_j|^2 + \mu \sum_{j=1}^k |y_k|^2 \geq  \frac{ M (z_k + 2 r)^2 }{ 1+ 2 \frac{m_a}{m_b} } .
\end{equation}
This concludes the proof.
\end{proof}

\begin{proof}[Proof of Lemma~\ref{supp_fj}.]
Suppose $(x^a,x^b,\tilde y) \in \supp f^r_{3,j}$.
If $j<k$, we need 
\begin{equation}
k\frac{\max\{x^a_{j+1},...,x^a_k\}-x^a_j}{2r}+\frac{3}{2}\leq 2
\end{equation}
for the factor $\chi_1$ to be non-zero.
This is equivalent to $\max\{x^a_{j+1},...,x^a_k\}\leq x^a_j+\frac{r}{k}$.
Thus, for any $1\leq j\leq k$ we have $\max\{x^a_{j},...,x^a_k\}\leq x^a_j+\frac{r}{k}$ on the support of $f^r_{3,j}$.
Let us argue inductively why $\max\{x^a_{1},...,x^a_k\}\leq x_j^a+r$.
Suppose we know for some $1<l\leq j$ that $\max\{x^a_{l},...,x^a_k\}\leq x_j^a+(j+1-l) \frac{r}{k}$.
If $x^a_{l-1} \leq \max\{x^a_{l},...,x^a_k\}$, we trivially have $\max\{x^a_{l-1},...,x^a_k\}\leq x_j^a+(j+1-(l-1)) \frac{r}{k}$.
If $x^a_{l-1} > \max\{x^a_{l},...,x^a_k\}$, for the factor $\chi_2\left(k\frac{\max\{x^a_{l},...,x^a_k\}-x^a_{l-1}}{2r}+\frac{3}{2}\right)$ not to vanish we have $\max\{x^a_{l},...,x^a_k\}+\frac{r}{k}\geq x^a_{l-1}$.
Thus, $\max\{x^a_{l-1},...,x^a_k\}=x^a_{l-1}\leq \max\{x^a_{l},...,x^a_k\}+\frac{r}{k}\leq x_j^a+(j+1-(l-1)) \frac{r}{k}$.
Inductively, we see that for every $j$ we have $\max\{x^a_{1},...,x^a_k\}\leq x_j^a+j \frac{r}{k} \leq x_j^a+r $.
Thus, $\supp f_{3,j} \cap \Omega_2 \subset \Omega_{3,j}$.

For the support of $\nabla f_{3,j}$, we have
\begin{equation}
\supp \nabla f^r_{3,j} \cap \Omega_2\subset \supp f^r_{3,j}\cap \Omega_2\subset \Omega_{3,j}=\{(x^a,x^b,\tilde y)\in \Omega_2 \vert x^a_j \geq \max\{x^a_1,...,\widehat{ x^a_j},...,x^a_k\} - r\}.
\end{equation}
Now, suppose $ x^a_j > \max\{x_1^a,...,\widehat{ x^a_j},...,x^a_k\}+r$.
It is sufficient to show that $f^r_{3,j}\equiv 1$ in this region.
For $j<k$, we have
\begin{equation}
k\frac{\max\{x^a_{j+1},...,x^a_k\}-x^a_j}{2r}+\frac{3}{2}\leq k\frac{ \max\{x_1^a,...,\widehat{ x^a_j},...,x^a_k\}-x^a_j}{2r}+\frac{3}{2} < -\frac{k}{2}+\frac{3}{2} \leq 1.
\end{equation}
Thus, $\chi_1\left(k\frac{\max\{x^a_{j+1},...,x^a_k\}-x^a_j}{2r}+\frac{3}{2}\right)=1$.
For $l <j \leq k$, we have
\begin{equation}
k\frac{\max\{x^a_{l+1},...,x^a_k\}-x^a_l}{2r}+\frac{3}{2}= k\frac{ x^a_j-x^a_l}{2r}+\frac{3}{2} \geq k\frac{x^a_j-\max\{x^a_1,...,\widehat{ x^a_j},...,x^a_k\}}{2r}+\frac{3}{2}> \frac{k}{2}+\frac{3}{2}\geq 2.
\end{equation}
Thus, $\chi_2\left(k\frac{\max\{x^a_{l+1},...,x^a_k\}-x^a_l}{2r}+\frac{3}{2}\right)=1$.
In total, $f_{3,j}\equiv1$ for $ x^a_j > \max\{x_1^a,...,\widehat{ x^a_j},...,x^a_k\}+r$.
\end{proof}

\bigskip
\noindent \textbf{Acknowledgments:} We thank Rupert Frank for contributing Appendix \ref{sec:app_b}. Funding from the European Union’s Horizon 2020 research and innovation programme under the ERC grant agreement No.~694227 is gratefully acknowledged. 

\appendix
\section{Appendix}\label{sec:app}

\subsection{Explicit example in one dimension}\label{sec:ex}

To illustrate the effect of a boundary on two-particle bound states, we present an explicit example in one dimension.
We consider particles with equal masses $m_a=m_b=\frac{1}{2}$ and with delta-interaction $V(y)=-\alpha \delta(y)$ for $\alpha>0$.
The full Hamiltonian is 
\begin{equation}
H=-\left(\frac{\partial}{\partial x^a}\right)^2-\left(\frac{\partial}{\partial x^b}\right)^2-\alpha \delta(x^a-x^b),
\end{equation}
either on $L^2(\BR^2)$ or on $L^2((0,\infty)^2)$ with Neumann boundary conditions.
In the first case, corresponding to $k=0$, we look at the operator $H_0=-2\frac{\partial^2}{\partial y^2}-\alpha \delta(y)$ on $L^2(\BR)$.
It has the ground state $\psi_0(y)=e^{-\frac{\alpha}{4}\vert y\vert}$
with corresponding energy $E^0=-\frac{\alpha^2}{8}$.

The second case corresponds to $k=1$.
To compute the ground state of $H=H_1$ on $L^2((0,\infty)^2)$, we mirror the problem along the $x^a=0$ and $x^b=0$ boundaries, and 
 look for the ground state of the modified Hamiltonian
\begin{equation}
\widetilde H_1=-\left(\frac{\partial}{\partial x^a}\right)^2-\left(\frac{\partial}{\partial x^b}\right)^2-\alpha \delta(x^a-x^b)-\alpha \delta(x^a+x^b)
\end{equation}
on $L^2(\BR^2)$.
This is exactly the operator considered in Proposition~\ref{mirror_details}.
Switching to relative and center of mass coordinates $y=x^a-x^b$ and $z=\frac{x^a+x^b}{2}$, we obtain
\begin{equation}
\widetilde H_1=\left(-2\frac{\partial^2}{\partial y^2}-\alpha \delta(y)\right)+\frac{1}{2}\left(-\frac{\partial^2}{\partial z^2}-\alpha \delta(z)\right).
\end{equation}
The ground state of $\tilde H_1$ is $\widetilde \psi_1 (y,z)=\psi_0(y) e^{-\frac{\alpha}{2} \vert z\vert}$, which decays exponentially away from the Neumann boundary.
The ground state energy $E^1=-\frac{\alpha^2}{4}$ is strictly lower than $E^0$.

\subsection{Proof of Lemma~\ref{h_l^N}}\label{sec:pf_lea_hlN}
Let $1\leq k \leq d$.
First, we shall prove that the claim is true for $l=1$, i.e.~$\lim_{L\to \infty} \inf \sigma(H_{k-1}^{L,1})\geq E^{k-1}$.
In $\Omega_{k-1}^{L,1}$, the first component of $y$ is constrained to $\vert y_1\vert<L$. 
Apart from that, $\Omega_{k-1}^{L,1}$ is the same as $Q_{k-1}\times \BR^{d-k+1}$ with components $1$ and $k$ swapped.
We localize in the $y_1$ direction, analogously to the one-dimensional case in Proposition A.5~in \cite{egger_bound_2020}.
For this, let $\chi_1,\chi_2:\BR\to [0,1]$ be continuously differentiable functions satisfying $\chi_1(t)=0$ for $t\geq 1$, $\chi_1(t)=1$ for $t\leq \frac{1}{2}$, and $\chi_1(t)^2+\chi_2(t)^2=1$ for all $t$.
Note that $c:=\max\{\lVert( \chi'_1)^2 \rVert_\infty,\lVert( \chi'_2)^2 \rVert_\infty\}< \infty $.
We choose the localizing functions $f_j$ on $\Omega_{k-1}^{L,1}$ as $f_j(z_2,\dots z_k,y)=\chi_j(\vert y_1\vert /L)$.
By the IMS localization formula, we have for all $\psi\in H^1(\Omega_{k-1}^{L,1})$
\begin{equation}
h_{k-1}^{L,1}[\psi]=h_{k-1}^{L,1}[f_1 \psi]+h_{k-1}^{L,1}[f_2 \psi]-\frac{1}{2\mu}\int_{\Omega_{k-1}^{L,1}}\left((\nabla f_1)^2+(\nabla f_2)^2\right)\vert \psi \vert^2.
\end{equation}
Note that $(\nabla f_j)^2=\frac{1}{L^2}(\chi_j'(\vert y_1\vert /L) )^2\leq \frac{c}{L^2}$.
Since $f_2\psi$ is nonzero only for $\vert y_1 \vert>L/2$, for large enough $L$, we have $h_{k-1}^{L,1}[f_2 \psi] \geq E^{k-1} \lVert f_2 \psi \rVert_2^2$ by Assumption~\ref{assump_V_general}\eqref{Vto0}.
Furthermore, since $f_1 \psi$ satisfies Dirichlet boundary conditions at $\vert y_1 \vert = L$, we can extend the function by zero to $y_1\in \BR$.
Additionally, let us swap the first and the $k$th components and call the function obtained this way $\iota(f_1 \psi)$. 
Note that $\iota(f_1 \psi)\in H^1(Q_{k-1} \times \BR^{d-k+1})$ and $\lVert  \iota(f_1 \psi) \rVert_2^2= \lVert f_1 \psi \rVert_2^2$.
Therefore,
\begin{equation}
\frac{h_{k-1}^{L,1}[f_1 \psi]}{\lVert f_1 \psi \rVert_2^2}=\frac{h_{k-1}[\iota(f_1 \psi)]}{\lVert  \iota(f_1 \psi) \rVert_2^2}\geq E^{k-1}.
\end{equation}
Combining the estimates, we obtain for large $L$ that 
\begin{equation}
\frac{h_{k-1}^{L,1}[\psi]}{\lVert \psi \rVert^2}\geq E^{k-1}\frac{ \lVert f_1\psi \rVert^2+\lVert f_2\psi \rVert^2}{\lVert \psi \rVert_2^2}-\frac{c}{\mu L^2}=E^{k-1}-\frac{c}{\mu L^2}.
\end{equation}
Hence, $ \inf \sigma(H_{k-1}^{L,1})\geq E^{k-1}-\frac{c}{\mu L^2}$ and the claim follows.

Note that for $k=1$, $l=1$ was the only possible case.
Consider $k \geq 2$.
We proceed by induction.
For $l\geq 2$, assume the claim holds for $l-1$.
The strategy is to bound $h_{k-1}^{L,l}$ using $h_{k-1}^{L,l-1}$ and $h_{k-2}^{L,l-1}$.
In $\Omega_{k-1}^{L,l}$, each of the first $l-1$ components are restricted to the (red) triangular domain 2 in Figure~\ref{essspec_domains}.
Furthermore, $y_l\in (-L,L)$ while in the $z$-coordinate the $l$th component is omitted.
In the components $l+1$ to $k$ we have the full quadrant.
Recall that $\delta=M/\max\{m_a,m_b\}$.
In the $(l-1)$th component, we localize such that one function has Dirichlet boundary conditions along the (red) line $z_{l-1}=L/\delta$ in Figure~\ref{essspec_domains} and the other is localized at $L/2\delta<z_{l-1}<L/\delta$, with a Dirichlet boundary at $z_{l-1}=L/2\delta$.
For this, we use the functions $f_j(z_1, \dots ,\hat z_l, \dots , z_k, y)=\chi_j(\delta  z_{l-1} /L)$.
By the IMS localization formula, we have for all $\psi\in H^1(\Omega_{k-1}^{L,l})$
\begin{equation}
h_{k-1}^{L,l}[\psi]=h_{k-1}^{L,l}[f_1 \psi]+h_{k-1}^{L,l}[f_2 \psi]-\frac{1}{2M}\int_{\Omega_{k-1}^{L,l}}\left((\nabla f_1)^2+(\nabla f_2)^2\right)\vert \psi \vert^2.
\end{equation}
Note that $(\nabla f_j)^2=\frac{\delta^2}{L^2}(\chi_j'(\delta z_{l-1} /L) )^2\leq \frac{\delta^2 c}{L^2}$.
Since $f_1\psi$ satisfies Dirichlet boundary conditions along $z_{l-1} =L/\delta$, one can extend the function by zero to the quadrant $Q_1$ in the $(l-1)$th component.
Additionally swap $y_{l-1}$ and $y_l$ to define $\iota_1(f_1 \psi)\in H^1(\Omega_{k-1}^{L,l-1})$. 
Then $\lVert  \iota_1(f_1 \psi) \rVert_2^2= \lVert f_1 \psi \rVert_2^2$ and hence
\begin{equation}
\frac{h_{k-1}^{L,l}[f_1 \psi]}{\lVert f_1 \psi \rVert_2^2}=\frac{h_{k-1}^{L,l-1}[\iota_1(f_1 \psi)]}{\lVert  \iota_1(f_1 \psi) \rVert_2^2}\geq \inf \sigma(H_{k-1}^{L,l-1}).
\end{equation}
To estimate $h_{k-1}^{L,l}[f_2 \psi]$, we localize in the $y_{l-1}$-direction, such that the first function satisfies Dirichlet boundary conditions at $y_{l-1}=L/2$ and the second function is nonzero only for $y_{l-1}>L/4$.
For this, we use the functions $g_j(z_1, \dots ,\hat z_l, \dots , z_k, y)=\chi_j(2 y_{l-1} /L)$.
The IMS localization formula gives
\begin{equation}
h_{k-1}^{L,l}[f_2\psi]=h_{k-1}^{L,l}[g_1 f_2 \psi]+h_{k-1}^{L,l}[g_2 f_2 \psi]-\frac{1}{2\mu}\int_{\Omega_{k-1}^{L,l}}\left((\nabla g_1)^2+(\nabla g_2)^2\right)\vert f_1 \psi \vert^2,
\end{equation}
where $(\nabla g_j)^2=\frac{4}{L^2}(\chi_j'(2 \vert y_{l-1}\vert /L) )^2\leq \frac{4 c}{L^2}$.
For $L$ large enough, by Assumption~\ref{assump_V_general}\eqref{Vto0}, we have $h_{k-1}^{L,l}[g_2 f_2 \psi] \geq E^{k-1} \lVert  g_2 f_2 \psi \rVert_2^2$.
In the $(l-1)$th component, the function $g_1 f_2 \psi$ is supported in the parallelogram $(z_{l-1},y_{l-1})\in (L/2\delta, L/\delta)\times (-L/2,L/2)$ and satisfies Dirichlet boundary conditions at $\vert y_{l-1}\vert=L/2$ and $z_{l-1}=L/2\delta$.
We extend the function $g_1 f_2 \psi$ by zero to $y_{l-1}\in \BR$. 
Then we define $\iota_2(g_1f_2\psi)$ on $\Omega_{k-2}^{L,l-1}\times (L/2\delta, L/\delta)$ as $\iota_2(g_1f_2\psi)(z_1\dots, \hat z_{l-1},\dots z_{k-1}, y, x)=g_1f_2\psi(z_1,\dots, z_{l-2}, x, z_l, \dots ,z_{k-1},y_1,\dots, y_{l-2},y_k, y_{l-1}, \dots, y_{k-1},y_{k+1}, \dots y_d )$.
Observe that $h_{k-1}^{L,l}$ now can effectively be decomposed into $h_{k-2}^{L,l-1}$ plus a Laplacian in the $x$-direction
\begin{equation}
\frac{h_{k-1}^{L,l}[g_1 f_2 \psi]}{\lVert g_1 f_2 \psi\rVert_2^2}=\frac{(h_{k-2}^{L,l-1}\otimes \BI+\BI\otimes q)[\iota_2(g_1 f_2 \psi)]}{\lVert \iota_2(g_1 f_2 \psi)\rVert_2^2},
\end{equation}
where $q$ is defined on $H^1((L/2\delta, L/\delta))$ through
\begin{equation}
q[\varphi]=\int_{L/2\delta}^{L/\delta} \frac{1}{2M} \vert \varphi' (x)\vert^2 \dd x.
\end{equation}
Since $\inf \sigma (H_{k-2}^{L, l-1} \otimes \BI - \frac{1}{2M}\BI\otimes \Delta_x) \geq \inf \sigma (H_{k-2}^{L, l-1} )$, we obtain
\begin{equation}
\frac{h_{k-1}^{L,l}[g_1 f_2 \psi]}{\lVert g_1 f_2 \psi\rVert_2^2}\geq \inf \sigma (H_{k-2}^{L, l-1} ).
\end{equation}
Combining all the estimates, we obtain that for large $L$ and all $\psi \in H^1(\Omega_{k-1}^{L,l})$
\begin{equation}
\frac{h_{k-1}^{L,l}[\psi]}{\lVert \psi \rVert^2}\geq \min\{\inf \sigma(H_{k-1}^{L,l-1}), \inf \sigma(H_{k-2}^{L,l-1}), E^{k-1}\}-\frac{\delta^2 c}{M L^2}-\frac{4c}{\mu L^2}.
\end{equation}
Taking $L\to \infty$ the claim now follows from the induction hypothesis.

\subsection{Technical details}\label{sec:tech-det}

By mirroring along the $x_j^a=0$ and $x_j^b=0$ hyperplanes, we can relate $H_k$ to an operator $\widetilde H_k$  defined in $L^2(\BR^{d+k})$. 

\begin{thm6}\label{mirror_details}
Let  $\widetilde H_k$ be the operator defined by the quadratic form 
\begin{equation}
\widetilde h_k [\psi]=\int_{ \BR^{d+k}}\left(\frac{1}{2m_a} \vert \nabla_{ x^a} \psi\vert^2+\frac{1}{2m_b} \vert \nabla_{ x^b} \psi\vert^2+\frac{1}{2\mu} \vert \nabla_{\tilde y} \psi\vert^2+V((\vert x^a_j\vert - \vert x^b_j \vert)_{j=1}^k,\tilde y) \vert \psi \vert^2\right) \dd x^a\dd x^b \dd \tilde y
\end{equation}
with domain ${D}[\widetilde h_k]=H^1(\BR^{d+k})$.
Then
$\inf \sigma(H_k)=\inf \sigma(\widetilde H_k)$ and $\inf \sigma_{\text{ess}}(H_k)=\inf \sigma_{\text{ess}}(\widetilde H_k)$.
Moreover, the function $\psi_{k}$ is a ground state of $H_k$ if and only if the function 
\begin{equation}
\widetilde \psi_k(x^a,x^b,\tilde y)=\psi_k((\vert x^a_j\vert)_{j=1}^k,(\vert x^b_j\vert)_{j=1}^k,\tilde y)
\end{equation} is a ground state of $\widetilde H_k$. 
\end{thm6}
\begin{proof}
The operator $\widetilde H_k$ commutes with all reflections along the $x^a_j=0$ or $x^b_j=0$ hyperplanes.
Reflections along different hyperplanes commute as well.
Therefore, the Hilbert space $\BH = L^2(\BR^{d+k})$ splits into subspaces $\BH=\bigoplus_r \BH_r$ characterized by the eigenvalues $\pm 1$ of these reflections.
We can write $\widetilde H_k = \bigoplus_r \widetilde H_k^r$, where $\widetilde H_k^r$ is the restriction of $\widetilde H_k$ to $\BH_r$.
For the spectrum, we obtain $\inf \sigma(\widetilde H_k)=\min_r \inf \sigma(\widetilde H_k^r)$ and $\inf \sigma_{\text{ess}}(\widetilde H_k)=\min_r\inf \sigma_{\text{ess}}(\widetilde H_k^r)$.

The subspace that is symmetric under all reflections corresponds to Neumann boundary conditions on $[0,\infty)^{2k} \times \BR^{d-k}$.
The other subspaces $\BH_r$ are antisymmetric under at least one reflection, so they have Dirichlet boundary conditions along the corresponding hyperplane.
Thus, the domains of the quadratic forms for $\widetilde H_k^r$ satisfy $D[h_k^r]\subset D[h_k^{\rm sym}]$.
By the min-max principle, $E_n( \widetilde H_k^r) \geq E_n(\widetilde H_k^{\rm sym})$.
Therefore, both  $\inf\sigma(\widetilde H_k)=\inf\sigma(\widetilde H_k^{\rm sym})$ and $\inf\sigma_{\text{ess}}(\widetilde H_k)=\inf\sigma_{\text{ess}}(H_k^{\rm sym})$.

Note that the map $U: L^2([0,\infty)^{2k}\times \BR^{d-k})\to L^2_{\rm sym}(\BR^{d+k})$ that maps $\psi$ to $\widetilde \psi(x^a,x^b,\tilde y)=\frac{1}{2^k}\psi((\vert x^a_j\vert)_j,(\vert x^b_j\vert)_j,\tilde y)$ is unitary.
Since $\widetilde H_k^{\rm sym} = U H_k U^{-1}$, the operators are unitarily equivalent and $\sigma (\widetilde H_k^{\rm sym})= \sigma(H_k)$.
\end{proof}

The next lemma follows from the Sobolev inequality, see e.g.~Sections 8.8 and 11.3 in \cite{lieb_analysis_2001}.
\begin{thm2}\label{bound_cone}
Let $\Omega\subset \BR^d$ be a domain satisfying the cone property (as defined in \cite{lieb_analysis_2001}) with radius $R$ and opening angle $\theta$.
Let $V$ satisfy Assumption~\ref{assump_V_general}(\ref{V_reg}).
Then, for any $0<a<1$ there is a constant $b\in \BR$ (depending only on $d, R, \theta, V$ and $a$) such that
\begin{equation}
  \int_{\Omega} \vert V\vert \vert f\vert^2 \leq  a \lVert \nabla f\rVert_{L^2(\Omega)}^2+b \lVert f \rVert_{L^2(\Omega)}^2,
\end{equation}
 for all $f \in H^1(\Omega)$.
\end{thm2}

\begin{thm6}\label{H_self-adj}
Let $0\leq k \leq d$. Assumption~\ref{assump_V_general}(\ref{V_reg}) implies that in the quadratic form $h_k$ in \eqref{def:hk} the interaction term is infinitesimally form bounded with respect to the kinetic energy. By the KLMN theorem, there is a unique self-adjoint operator $H_k$ corresponding to $h_k$, and both $h_k$ and $H_k$ are bounded from below.
\end{thm6}

\begin{proof}
The quadratic form $q_k: H^1([0,\infty)^{2k}\times \BR^{d-k})\to \BR$ given by
\begin{equation}
q_k [\psi]=\int_{[0,\infty)^{2k}\times \BR^{d-k}} \left(\frac{1}{2m_a} \vert \nabla_{x^a} \psi\vert^2+\frac{1}{2m_b} \vert \nabla_{x^b} \psi\vert^2+\frac{1}{2\mu} \vert \nabla_{y} \psi\vert^2\right)\dd x^a \dd x^b \dd \tilde y
\end{equation}
is closed and bounded from below.
In order to apply the KLMN theorem, we need to show that there are constants $a<1, b\in \BR$ such that for all $\psi\in H^1([0,\infty)^{2k}\times \BR^{d-k})$ 
\begin{equation}
K[\psi]:=\left\vert \int_{[0,\infty)^{2k}\times \BR^{d-k}} V( x^a-x^b,\tilde y) \vert \psi \vert^2 \dd x^a\dd x^b \dd \tilde y\right \vert \leq a q_k[\psi]+b \lVert \psi \rVert^2_2.
\end{equation}
Let $\psi\in H^1([0,\infty)^{2k}\times \BR^{d-k})$ and define $\widetilde \psi(x^a,x^b,\tilde y)=\frac{1}{2^k}\psi((\vert x^a_j\vert)_j,(\vert x^b_j\vert)_j,\tilde y)$ for $(x^a,x^b,\tilde y) \in\BR^k\times \BR^k \times \BR^{d-k}$.
We have $\lVert \widetilde  \psi \rVert^2_2= \lVert \psi \rVert^2_2$ and $\lVert \nabla \widetilde \psi \rVert^2_2= \lVert \nabla\psi \rVert^2_2$.
Moreover, $\psi$ and $2^k \widetilde \psi$ agree on $[0,\infty)^{2k}\times \BR^{d-k}$. 
Hence, 
\begin{equation}\label{a19b}
K[\psi] 
\leq  4^k\int_{[0,\infty)^{2k}\times \BR^{d-k}} \vert V(x^a-x^b,\tilde y)\vert  \vert \widetilde \psi(x^a,x^b,\tilde y)\vert^2 \dd x^a \dd x^b \dd \tilde y.
\end{equation}
Since the integrand is nonnegative, extending the domain of integration from $[0,\infty)^{2k}\times \BR^{d-k}$ to $\BR^{2k}\times \BR^{d-k}$ gives the upper bound
\begin{multline}\label{a20b}
K[\psi] 
\leq 4^k \int_{\BR^{2k}\times \BR^{d-k}} \vert V(x^a-x^b,\tilde y)\vert  \vert \widetilde \psi(x^a,x^b,\tilde y)\vert^2 \dd x^a \dd x^b \dd \tilde y \\
=4^k \int_{\BR^{k}\times \BR^{d}} \vert V(y)\vert \vert \widetilde \psi(w+(y_1,...,y_k)/2,w-(y_1,...,y_k)/2,\tilde y)\vert^2 \dd w \dd y,
\end{multline}
where we changed to coordinates $w=\frac{x^a+x^b}{2}$ and $y$.
For almost every $w\in \BR^k$, the function $f(y)= \widetilde \psi(w+(y_1,...,y_k)/2,w-(y_1,...,y_k)/2,\tilde y)$ lies in $H^1(\BR^d)$ by Fubini's theorem.
By Lemma~\ref{bound_cone}, for any $0<\tilde a$ there is a constant $b$ independent of $f$ such that $\int_{\BR^d} \vert V\vert \vert f\vert^2 \leq  \tilde a \lVert \nabla f\rVert_2^2+b \lVert f \rVert_2^2$.
Integrating over $w$ then gives
\begin{equation}\label{a10}
K[\psi] \leq 4^{k}\left( \tilde a \int_{\BR^{k}\times \BR^{d}} \left\vert \nabla_y\widetilde \psi(w+(y_1,...,y_k)/2,w-(y_1,...,y_k)/2,\tilde y)\right\vert^2 \dd w \dd y+ b\lVert \widetilde \psi \rVert_2^2\right).
\end{equation}
For $1\leq j\leq k$, 
\begin{equation}\label{a11}
\left \vert \partial_{y_j} \widetilde \psi(w+(y_1,...,y_k)/2,w-(y_1,...,y_k)/2,\tilde y)\right\vert^2=\frac{1}{4}\left\vert \partial_{x^a_j}\widetilde\psi-\partial_{x^b_j}\widetilde \psi\right\vert^2 \leq \frac{1}{2} \left(\left\vert \partial_{x^a_j}\widetilde\psi\right\vert^2+\left\vert \partial_{x^b_j}\widetilde\psi\right\vert^2\right).
\end{equation}
Therefore,
\begin{equation}\label{a12}
K[\psi] \leq 4^{k}\left( \tilde a \lVert \nabla\widetilde \psi \rVert_2^2+ b\lVert \widetilde \psi \rVert_2^2\right)
= 4^k \tilde a \lVert \nabla \psi \rVert_2^2+ 4^k b \lVert \psi \rVert_2^2.
\end{equation}
For any $0< a <1$, pick $\tilde a=2^{-2k-1}\min{(m_a^{-1},m_b^{-1})} a$ to obtain $K[\psi]\leq a q_k[\psi]+ 4^k b \lVert \psi \rVert_2^2.$
\end{proof}

\begin{thm2}\label{a_self-adj_prop_1}
The quadratic forms defined in the proof of Proposition~\ref{esspec_general} in Eqs.~\eqref{al_prop1} and \eqref{h_l^N_def} correspond to unique self-adjoint operators.
\end{thm2}
\begin{proof}
In all cases we prove that the potential term in the quadratic form is infinitesimally bounded with respect to the kinetic energy term.
The claim then follows from the KLMN theorem.

Let us begin with the quadratic form $h_{k-1}^{l,L}$ in \eqref{h_l^N_def}.
The idea is to use the same mirroring argument as in Prop.~\ref{H_self-adj} for the coordinate components from $l+1$ to $k$.
In the first $l-1$ components, we extend the triangular domain in Figure~\ref{essspec_domains} via a suitable mirroring, in order to be able to apply Lemma~\ref{bound_cone}.
To be precise, we define the map $\phi$ taking $(0,L/\delta)\times \left(-\frac{M L}{m_b \delta},\frac{M L}{m_a \delta}\right)$ to the triangular domain $\{(z,y)\in (0,L/\delta)\times \BR \vert- \frac{M}{m_b} z < y < \frac{M}{m_a} z \}$ as
\begin{align}
\phi(z,y) &= (z,y) \quad &\text{if} \ x^a=z+\frac{m_b}{M} y\geq0\ \text{and} \ x^b=z-\frac{m_a}{M} y\geq0\\
\phi(z,y) &= \left(\frac{m_a}{M}y,\frac{M}{m_a} z\right) &\text{if} \ x^b \leq0\\
\phi(z,y) &= \left(\frac{m_b}{M}y,\frac{M}{m_b} z\right) &\text{if} \ x^a \leq0
\end{align}
Let us use the notation $\phi=(\phi_1,\phi_2)$.
Note that for a function $f$ defined on the triangular domain, we have 
\begin{equation}\label{normeqphi}
\lVert f \circ \phi \rVert_2^2=
2 \lVert f \rVert_2^2,
\end{equation}
where one contribution of $\lVert f \rVert_2^2$ comes from the triangular domain, and the second $\lVert f \rVert_2^2$ is the sum of the contributions with $x^b<0$ and $x^a<0$.
In the region with $x^b<0$ we have
\begin{multline}
\int_{0}^{\frac{M L}{m_a \delta}}\dd y \int_0^{\frac{m_a y}{M }}\dd z  \vert f(\phi(z,y))\vert^2=\int_{0}^{\frac{M L}{m_a \delta}}\dd y \int_0^{\frac{m_a y}{M }}\dd z  \vert f(m_a y/M, M z/m_a)\vert^2\\
=\int_0^{L/\delta} \dd \tilde z \int_0^{\frac{M \tilde z}{m_a }} \dd \tilde y  \vert f(\tilde z, \tilde y)\vert^2,
\end{multline}
where we substituted $\tilde z= m_a y/M $ and $\tilde y = M z/m_a$.
Similarly, for $x^a<0$
\begin{equation}
\int_{-\frac{M L}{m_b \delta}}^0\dd y \int_0^{\frac{m_b y}{M }}\dd z  \vert f(\phi(z,y))\vert^2
=\int_0^{L/\delta} \dd \tilde z \int_{-\frac{M \tilde z}{m_b }}^0 \dd \tilde y  \vert f(\tilde z, \tilde y)\vert^2.
\end{equation}
Moreover, if $f\in H^1$, then $f\circ \phi \in H^1$ by the Lipschitz continuity of $\phi$.

Let us work in center of mass and relative coordinates in the first $l$ components, and with the $x^a$ and $x^b$ coordinates in components $l+1$ to $k$.
The kinetic part of $h_{k-1}^{l,L}$ is then
\begin{multline}
q_{k-1}^{l,L}[\psi]:=\int_{\Omega_{k-1}^{l,L}}\left[ \rule{0cm}{0.8cm}\right.\sum_{j=1}^{l-1} \left( \frac{1}{2M}\vert \nabla_{z_j} \psi \vert^2+\frac{1}{2\mu}\vert \nabla_{y_j} \psi \vert^2\right)+\frac{1}{2\mu}\vert \nabla_{y_l} \psi \vert^2+ \sum_{j=l+1}^k\left( \frac{1}{2m_a}\vert \nabla_{x^a_j} \psi \vert^2 \right.\\
\left. +\frac{1}{2m_b}\vert \nabla_{x^b_j} \psi \vert^2\right)+\frac{1}{2\mu}\vert \nabla_{\tilde y} \psi \vert^2 \left. \rule{0cm}{0.8cm}\right]\dd z_1 \dots \dd z_{l-1} \dd x^a_{l+1} \dots \dd x^a_k \dd y_1 \dots \dd y_l \dd x^b_{l+1} \dots \dd x^b_k \dd \tilde y.
\end{multline}
For $\psi\in H^1(\Omega_{k-1}^{l,L})$ define $\widetilde \psi $ on 
\begin{multline} 
\widetilde \Omega_{k-1}^{l,L}:=\left \{ (z_1,\dots z_{l-1},x^a_{l+1},\dots x^a_{k},y_1, \dots ,y_l, x^b_{l+1},\dots x^b_{k},\tilde y) \vert \forall j<l: z_j \in (0,L/\delta), \right. \\
\left.  y_j   \in \left(-\frac{M L}{m_b \delta},\frac{M L}{m_a \delta}\right), y_l\in (-L,L) , \forall l<j\leq k: x^a_j \in \BR, x^b_j \in \BR ,\tilde y \in \BR^{d-k} \right \}
\end{multline}
as
\begin{equation}
\widetilde \psi(z,y)=\frac{1}{2^{(l-1)/2}}\frac{1}{2^{k-l}}\psi\left((\phi_1(z_j,y_j))_{j=1}^{l-1},( \vert x^a_j\vert)_{j=l+1}^k,(\phi_2(z_j,y_j))_{j=1}^{l-1},y_l,(\vert x^b_j\vert)_{j=l+1}^{k},\tilde y\right).
\end{equation}
By \eqref{normeqphi} we have $\lVert \widetilde  \psi \rVert^2_2= \lVert \psi \rVert^2_2$.
Furthermore, $\lVert \nabla \widetilde \psi \rVert^2_2 \leq \left(\frac{M^2}{\min\{m_a,m_b\}^2}+1\right)^{l-1}\lVert \nabla\psi \rVert^2_2$.

Analogously to \eqref{a19b}-\eqref{a20b} we obtain
\begin{multline}\label{a33b}
K[\psi]:=\\
\left\vert \int_{\Omega_{k-1}^{l,L}}V(y_1, \dots y_l, x^a_{l+1}-x^b_{l+1}, \dots x^a_k-x^b_k,\tilde y)\vert \psi \vert^2\dd z_1 \dots \dd z_{l-1} \dd x^a_{l+1} \dots \dd x^a_k \dd y_1 \dots \dd y_l \dd x^b_{l+1} \dots \dd x^b_k \dd \tilde y\right \vert \\
\leq 2^{l-1} 4^{k-l} \int_{\widetilde \Omega_{k-1}^{l,L}}\vert  V(y)\vert \vert \widetilde \psi(z_1,...z_{l-1},(w_j+\frac{y_j}{2})_{j=l+1}^k, y_1,...,y_l, (w_j-\frac{y_j}{2})_{j=l+1}^k,\tilde y) \vert^2 \\
\dd z_1\dots \dd z_{l-1} \dd w_{l+1} \dots \dd w_k \dd y ,
\end{multline}
where we changed the coordinates $x^a_j,x^b_j$ to $w_j=\frac{x^a+x^b}{2}$ and $y_j$.
Let $D_y=\left(-\frac{M L}{m_b \delta},\frac{M L}{m_a \delta}\right)^{l-1}\times (-L,L)\times\BR^{d-k}$.
For almost every $(z_1, \dots z_{l-1},w_{l+1}, \dots w_k)\in (0,L/\delta)^{l-1}\times \BR^{k-l}$, the function $f(y)= \widetilde \psi(z_1,...z_{l-1},(w_j+\frac{y_j}{2})_{j=l+1}^k, y_1,...,y_l, (w_j-\frac{y_j}{2})_{j=l+1}^k,\tilde y) $ lies in $H^1\left(D_y\right)$ by Fubini's theorem.
Applying Lemma~\ref{bound_cone} with $\Omega=D_y$ and integrating over $z$ and $w$ one obtains
\begin{multline}
K[\psi] \leq 2^{l-1} 4^{k-l} a \int_{\tilde \Omega_{k-1}^{l,L}} \left\vert \nabla_y\widetilde \psi(z_1,...z_{l-1},(w_j+\frac{y_j}{2})_{j=l+1}^k, y_1,...,y_l, (w_j-\frac{y_j}{2})_{j=l+1}^k,\tilde y) \right\vert^2\\
 \dd z_1 \dots \dd z_{l-1} \dd w_{l+1} \dots \dd w_k \dd y
+ 2^{l-1} 4^{k-l}b\lVert \widetilde \psi \rVert_2^2
\end{multline}
for any $a>0$ and a suitable constant $b$.
As in \eqref{a11} we have
\begin{multline}\label{a35b}
K[\psi] \leq 2^{l-1} 4^{k-l}\left( a \lVert \nabla\widetilde \psi \rVert_2^2+ b\lVert \widetilde \psi \rVert_2^2\right) \\
\leq 2^{l-1} 4^{k-l}  \left(\frac{M^2}{\min\{m_a,m_b\}^2}+1\right)^{l-1}  a \lVert \nabla \psi \rVert_2^2+2^{l-1} 4^{k-l} b \lVert \psi \rVert_2^2.
\end{multline}
Since $a$ can be arbitrarily small, the interaction term is infinitesimally bounded w.r.t. $q_{k-1}^{l,L}$.

Let us now consider the quadratic form $a_l$ in \eqref{al_prop1}.
For $l=k+2$, the potential term is bounded from below since $\vert y\vert >L$, and is hence infinitesimally bounded w.r.t the kinetic energy.

The kinetic part of $a_l$ is
\begin{multline}
q_l[\psi]:=\int_{\Omega_l}\left[ \rule{0cm}{0.8cm}\right.\sum_{j=1}^l \left( \frac{1}{2M}\vert \nabla_{z_j} \psi \vert^2+\frac{1}{2\mu}\vert \nabla_{y_j} \psi \vert^2\right)+ \sum_{j=l+1}^k\left( \frac{1}{2m_a}\vert \nabla_{x^a_j} \psi \vert^2+\frac{1}{2m_b}\vert \nabla_{x^b_j} \psi \vert^2\right)\\
+\frac{1}{2\mu}\vert \nabla_{\tilde y} \psi \vert^2 \left. \rule{0cm}{0.8cm}\right]\dd z_1 \dots \dd z_l \dd x^a_{l+1} \dots \dd x^a_k \dd y_1 \dots \dd y_l \dd x^b_{l+1} \dots \dd x^b_k \dd \tilde y.
\end{multline}
First, we consider $1\leq l\leq k$.
Then, $a_l$ is closely related to $h_{k-1}^{l,L}$ through \eqref{al=hl+q}.
Let $\psi \in H^1(\Omega_l)$.
For every $z_l\in(L/\delta,\infty)$, the function $\psi(\cdot,\dots , z_l, \dots,\cdot)$ belongs to $H^1(\Omega_{k-1}^{l,L})$.
In \eqref{a33b}-\eqref{a35b}, we saw that for any $a>0$ there is a constant $b$ such that
\begin{equation}
\int_{\Omega_{k-1}^{l,L}}\left\vert V(y) \right\vert \vert \psi(z,y)\vert^2  \dd y \dd z_1 \dots \widehat{\dd z_l} \dots \dd z_k   \leq a q_{k-1}^{l,L}[\psi(\cdot, z_l, \cdot)]+b \int \vert \psi(z,y)\vert^2  \dd y \dd z_1 \dots \widehat{\dd z_l} \dots \dd z_k .
\end{equation}
Integrating the inequality over $z_l$, we obtain
\begin{equation}
\int_{\Omega_l}\vert V(y)\vert \vert \psi(z,y)\vert^2 \dd y \dd z\leq a \int_{L/\delta}^\infty q_{k-1}^{l,L}[\psi(\cdot, z_l, \cdot)]\dd z_l+b \lVert \psi\rVert_2^2 \leq a q_l[\psi]+b \lVert \psi\rVert_2^2.
\end{equation}
Hence, the potential term is infinitesimally bounded w.r.t $q_l$.

For $l= k+1$, we use the map $\phi$ in the first $k$ components.
For $\psi\in H^1(\Omega_{k+1})$ define $\widetilde \psi $ on 
\begin{equation} 
\widetilde \Omega_{k+1}:=(0,L/\delta)^k\times \left(-\frac{M L}{m_b \delta},\frac{M L}{m_a \delta}\right)^k \times  (-L,L)^{d-k} 
\end{equation}
as
\begin{equation}
\widetilde \psi(z,y)=\frac{1}{{2}^{k/2}}\psi\left((\phi_1(z_j,y_j))_{j=1}^{k},(\phi_2(z_j,y_j))_{j=1}^{k},\tilde y\right).
\end{equation}
By \eqref{normeqphi} we have $\lVert \widetilde  \psi \rVert^2_2= \lVert \psi \rVert^2_2$.
Furthermore, $\lVert \nabla \widetilde \psi \rVert^2_2 \leq \left(\frac{M^2}{\min\{m_a,m_b\}^2}+1\right)^{k}\lVert \nabla\psi \rVert^2_2$.
Analogously to \eqref{a19b}-\eqref{a20b} we obtain
\begin{equation}
K[\psi]:=\left\vert \int_{\Omega_{k+1}}V(y)\vert \psi(z,y) \vert^2\dd z \dd y\right \vert 
\leq 2^{k}\int_{\widetilde \Omega_{k+1}}\vert  V(y)\vert \vert \widetilde \psi(z,y) \vert^2 \dd z \dd y.
\end{equation}
Let $D_y=\left(-\frac{M L}{m_b \delta},\frac{M L}{m_a \delta}\right)^{k}\times (-L,L)^{d-k}$.
For almost every $z\in (0,L/\delta)^{k}$, the function $f(y)= \widetilde \psi(z,y) $ lies in $H^1\left(D_y\right)$ by Fubini's theorem.
Applying Lemma~\ref{bound_cone} with $\Omega=D_y$ and integrating over $z$ gives
\begin{equation}
K[\psi] \leq 2^{k}  a \int_{\tilde \Omega_{k+1}} \left\vert \nabla_y\widetilde \psi(z, y) \right\vert^2 \dd z \dd y
+ 2^k b\lVert \widetilde \psi \rVert_2^2 \leq 2^{k}  a \lVert \nabla \widetilde \psi \rVert_2^2+ 2^k b\lVert \widetilde \psi \rVert_2^2 
\end{equation}
for any $a>0$ and a suitable constant $b$.
Hence, 
\begin{equation}
K[\psi] \leq 2^k \left(\frac{M^2}{\min\{m_a,m_b\}^2}+1\right)^{k}  a \lVert \nabla \psi \rVert_2^2+2^k b \lVert \psi \rVert_2^2.
\end{equation}
Since $a$ can be arbitrarily close to zero, the interaction term is infinitesimally bounded w.r.t. $q_{k+1}$.
\end{proof}

\begin{thm2}\label{a_self-adj_gen}
The quadratic forms defined in the proof of Theorem~\ref{Hk_fin_ev} in Eqs.~(\ref{aj_nr_ev_k}), (\ref{a_1_bullet}), (\ref{a_3_j}), (\ref{aj_4_5}), (\ref{a4_hat}), (\ref{b1}) and (\ref{b3})  correspond to unique self-adjoint operators.
\end{thm2}

\begin{proof}
The quadratic forms $a_j$ with $j\in \{1,2,4,5\}$ in Eqs.~(\ref{aj_nr_ev_k}) and (\ref{aj_4_5}) and the forms $a_{3,j}$ for $1\leq j\leq k$ in (\ref{a_3_j}) have the form
\begin{multline}
{a}_j[\varphi]= \int_{\Omega_j} \biggl(\frac{1}{2m_a} \vert \nabla_{x^a} \varphi\vert^2+ \frac{1}{2m_b}\vert \nabla_{x^b} \varphi\vert^2+\frac{1}{2\mu }  \vert \nabla_{\tilde y} \varphi\vert^2\\
+\left(V(x^a-x^b,\tilde y)+V_{\infty}(x^a,x^b,\tilde y)\right)  \vert \varphi \vert^2\biggl) \dd x^a \dd x^b \dd \tilde y
\end{multline}
for some bounded potential $V_{\infty}$.
The quadratic form $q_j: H^1(\Omega_j)\to \BR$ given by
\begin{equation}
q_j [\varphi]=\int_{\Omega_j} \left(\frac{1}{2m_a}\vert \nabla_{x^a} \varphi\vert^2+\frac{1}{2m_b} \vert \nabla_{x^b} \varphi\vert^2+\frac{1}{2\mu}  \vert \nabla_{\tilde y} \varphi\vert^2 \right)\dd x^a \dd x^b \dd \tilde y
\end{equation}
is closed and bounded from below.
Using that $\varphi \in D[a_j]$ vanishes outside $\overline{\Omega_j}$ and applying Proposition~\ref{H_self-adj}, we obtain
\begin{align}\nonumber
\left\vert \int_{\Omega_j}V(x^a-x^b,\tilde y)+V_{\infty}(x^a,x^b,\tilde y) \vert \varphi \vert^2 \right \vert & \leq \left\vert \int_{Q_k\times \BR^{d-k}} V( y) \vert \varphi \vert^2 \right \vert + \lVert V_{\infty}\rVert_{\infty}\lVert \varphi \rVert^2_2\\
& \leq a q_j[\varphi]+(b+ \lVert V_{\infty}\rVert_{\infty}) \lVert \varphi \rVert^2_2
\end{align}
for some $a<1$ and $b \in \BR$.
By the KLMN theorem, there is a unique self-adjoint operator $A_j$ corresponding to $a_j$.

For $\hat a_4$ in (\ref{a4_hat}), note that $K_R$ is bounded.
Adapting the argument in Proposition~\ref{H_self-adj}, we show that the interaction term is infinitesimally bounded with respect to
the kinetic part $\hat q: H^1(((0,\infty)^{k-1}\times \BR)^2 \times \BR^{d-k})\to \BR$ given by
\begin{equation}
\hat q[\varphi]= \int_{\hat \Omega_4} \left( \frac{1}{2m_a}\vert \nabla_{x^a} \varphi\vert^2+ \frac{1}{2m_b}\vert \nabla_{x^b} \varphi\vert^2+\frac{1}{2\mu}  \vert \nabla_{\tilde y} \varphi\vert^2\right) \dd x^a \dd x^b \dd \tilde y.
\end{equation}
For $\psi \in H^1(\hat \Omega_4)$, define define $\widetilde \psi(x^a,x^b,\tilde y)=\frac{1}{2^{k-1}}\psi((\vert x^a_j\vert)_{j=1}^{k-1}, x^a_k,(\vert x^b_j\vert)_{j=1}^{k-1}, x^b_k,\tilde y)$ for $(x^a,x^b,\tilde y) \in\BR^k\times \BR^k \times \BR^{d-k}$.
We have $\lVert \widetilde  \psi \rVert^2_2= \lVert \psi \rVert^2_2$ and $\lVert \nabla \widetilde \psi \rVert^2_2= \lVert \nabla\psi \rVert^2_2$.
Following the same steps as in Proposition~\ref{H_self-adj} from \eqref{a19b}-\eqref{a12} with this adapted choice of $\tilde \psi$, we obtain that for any $0< a$ there is a $b$ such that
\begin{equation}
K[\psi] :=\left\vert \int_{\hat \Omega_4} V( x^a-x^b,\tilde y) \vert \psi \vert^2 \dd x^a\dd x^b \dd \tilde y\right \vert \leq 4^{k-1}\left(  a \lVert \nabla\widetilde \psi \rVert_2^2+ b\lVert \widetilde \psi \rVert_2^2\right)
= 4^{k-1}  a \lVert \nabla \psi \rVert_2^2+ 4^k b \lVert \psi \rVert_2^2.
\end{equation}
By the KLMN theorem, $\hat a_4$ corresponds to a self-adjoint operator.
Since $b_1$ in (\ref{b1}) differs from $\hat a_4$ by a bounded term, it also corresponds to a self-adjoint operator.
For $b_2$ in (\ref{b3}) and $a_{1,\text{ext}}$ in (\ref{a_1_bullet}), the potential is bounded.
Thus, these forms also correspond to self-adjoint operators.

\begin{figure}[ht]
\begin{minipage}[t]{0.48\linewidth}
\centering
\begin{tikzpicture}[scale=0.3]

\draw[blue, very thick, decoration={markings, mark=at position 1 with {\arrow[scale=2,>=stealth]{>}}},
        postaction={decorate}]
 (0,0)--(10,10) ;
 \draw[blue, very thick, decoration={markings, mark=at position 1 with {\arrow[scale=2,>=stealth]{>}}},
        postaction={decorate}]
(0,0)--(-10,10);

\pattern[pattern color=gray!100, pattern=my north west lines]    (0,0) --++ (-5,5)-- ++ (0,5) -- ++(10,0) -- ++(0,-5);


\draw (0,0) node[below left]{$0$};
\draw (0,5) node[left]{$R$};
\draw (5,0) node[below]{$R$};
\draw (0,10) node[left]{$2w_j$};
\draw (10,0) node[below]{$y_j$};
\draw (10,10) node[right, blue]{$x^a_j$};
\draw (-10,10) node[left,blue]{$x^b_j$};
\draw (5,5) node[below right, blue]{$R$};

 \draw[black, very thick, decoration={markings, mark=at position 1 with {\arrow[scale=2,>=stealth]{>}}},
        postaction={decorate}]
 (-10,0)--(10,0);
 \draw[black, very thick, decoration={markings, mark=at position 1 with {\arrow[scale=2,>=stealth]{>}}},
        postaction={decorate}]
 (0,-10)--(0,10);

 \draw[black, very thick]
(-5,5)--(-5,10)
(5,5)--(5,10);

\draw[black, dashed](5,5)--(5,0) (0,5)--(5,5);

\end{tikzpicture} 
\caption{In the domain of $\psi$ for $1\leq j\leq k$, the coordinates $(x^a_j,x^b_j)$ lie in the hatched set. We have $y_j=x^a_j-x^b_j$ and $w_j=\frac{x^a_j+x^b_j}{2}$.}
\label{k2_mirror_1}
\end{minipage}
\hspace{0.02\linewidth}
\begin{minipage}[t]{0.48\linewidth}
\centering
\begin{tikzpicture}[scale=0.3]

  \draw[blue, very thick, decoration={markings, mark=at position 1 with {\arrow[scale=2,>=stealth]{>}}},
        postaction={decorate}]
 (-10,-10)--(10,10) ;
 \draw[blue, very thick, decoration={markings, mark=at position 1 with {\arrow[scale=2,>=stealth]{>}}},
        postaction={decorate}]
(10,-10)--(-10,10);

\pattern[pattern color=gray!100, pattern=my north west lines]    (-5,-10) --++ (0,5)-- ++ (-5,0) -- ++(0,10) -- ++(5,0)--++(0,5)--++(10,0)--++(0,-5)--++(5,0)--++(0,-10)--++(-5,0)--++(0,-5);


\draw (0,0) node[below left]{$0$};
\draw (0,5) node[left]{$R$};
\draw (5,0) node[below]{$R$};
\draw (0,10) node[left]{$2w_j$};
\draw (10,0) node[below]{$y_j$};
\draw (10,10) node[right, blue]{$x^a_j$};
\draw (-10,10) node[left,blue]{$x^b_j$};
\draw (5,5) node[below right, blue]{$R$};

 \draw[black, very thick, decoration={markings, mark=at position 1 with {\arrow[scale=2,>=stealth]{>}}},
        postaction={decorate}]
 (-10,0)--(10,0);
 \draw[black, very thick, decoration={markings, mark=at position 1 with {\arrow[scale=2,>=stealth]{>}}},
        postaction={decorate}]
 (0,-10)--(0,10);
 
 \draw[black, very thick]
(-5,5)--(-5,10) (5,5)--(5,10)
(-5,-5)--(-5,-10) (5,-5)--(5,-10)
(5,-5)--(10,-5) (5,5)--(10,5)
(-5,-5)--(-10,-5) (-5,5)--(-10,5);

\draw[black, dashed](5,5)--(5,0)  (0,5)--(5,5);

\end{tikzpicture} 
\caption{Mirroring $\psi$ along $x^a_j=0$ and $x^b_j=0$ defines $\tilde \psi$. For $1\leq j\leq k$, the coordinates $(x^a_j,x^b_j)$ or equivalently $(w_j,y_j)$ lie in the hatched set.}
\label{k2_mirror_2}
\end{minipage}
\end{figure}

For $a_{1,\text{int}}$ in (\ref{a_1_bullet}), we proceed similarly to Proposition~\ref{H_self-adj}.
Let $\psi\in D[a_{1,\text{int}}]$. 
The domain of $\psi$ is sketched in Figure~\ref{k2_mirror_1}.
Mirroring the domain along the $x^a_j=0$ and $ x^b_j=0$ hyperplanes, we obtain the set $\tilde \Omega$ sketched in Figure~\ref{k2_mirror_2}.
For $(x^a,x^b,\tilde y)\in \tilde \Omega$ define $\tilde \psi(x^a,x^b,\tilde y)= \frac{1}{2^k}\psi((\vert x^a_j\vert)_j,(\vert x^b_j\vert)_j,\tilde y)$.
We have $\lVert \tilde \psi \rVert^2_2= \lVert \psi \rVert^2_2$ and $\lVert \nabla \tilde \psi \rVert^2_2= \lVert \nabla\psi \rVert^2_2$.
Using the triangle inequality and enlarging the domain of integration to $\tilde \Omega$, we have
\begin{multline}
K[\psi]:=\left \vert \int_{\Omega_{1,\text{int}}} V(x^a-x^b,\tilde y)\vert \psi(x^a,x^b,\tilde y)\vert^2  \dd x^a \dd x^b \dd \tilde y \right \vert\\
\leq 4^k\int_{\tilde \Omega} \vert V(x^a-x^b,\tilde y)\vert \vert \tilde \psi(x^a,x^b,\tilde y)\vert^2  \dd x^a \dd x^b \dd \tilde y.
\end{multline}
We change to coordinates $w=\frac{x^a+x^b}{2}$ and $y$.
For every $w\in \BR^k$, the set 
\begin{equation}
\Omega_w=\{\ y\in \BR^d\vert (w+(y_1,...,y_k)/2,w-(y_1,...,y_k)/2,\tilde y) \in \tilde \Omega\}
\end{equation}
is equal to $I_1 \times ... \times I_k\times \BR^{d-k}$, where each $I_j\in \{\BR, (-R,R)\}$ (Figure~\ref{k2_mirror_2}).
Thus, there is an angle $\theta$ and radius $r$ such that all the sets $\Omega_w$ satisfy the cone property with parameters $\theta,r$.
For almost every $w\in \BR^k$, the function $f(y)=\widetilde \psi(w+(y_1,...,y_k)/2,w-(y_1,...,y_k)/2,\tilde y)$ lies in $H^1(\Omega_w)$.
By Lemma~\ref{bound_cone}, for any $0<\tilde a$ there is a constant $b$ independent of $f_w $ and $w$ such that 
\begin{equation}\label{a19}
\int_{\Omega_w} \vert V(y)\vert \vert f(y)\vert^2 \dd y \leq  \tilde a \lVert \nabla f\rVert_2^2+b \lVert f\rVert_2^2.
\end{equation}
Integrating inequality (\ref{a19}) over $w$ and using (\ref{a11}) gives 
\begin{equation}
\int_{\tilde \Omega} \vert V(x^a-x^b,\tilde y)\vert \vert \widetilde \psi(x^a,x^b,\tilde y) \vert^2 \dd x^a \dd x^b \dd \tilde y \leq  \tilde a \lVert \nabla_y \widetilde \psi \rVert^2+b \lVert \widetilde \psi \rVert_2^2\leq  \tilde a \lVert \nabla \widetilde \psi \rVert^2+b \lVert \widetilde \psi \rVert_2^2.
\end{equation}
In total, we thus have
\begin{equation}
K[\psi]  \leq 4^k \tilde a \lVert \nabla \psi \rVert_2^2+ 4^k b \lVert \psi \rVert_2^2.
\end{equation}
For any $0<a<1$, pick $\tilde a=2^{-2k-1}\min(m_a^{-1},m_b^{-1})a$ to obtain $K[\psi] \leq a q_{1,\text{int}}[\psi]+ 4^k b \lVert \psi \rVert_2^2.$
The KLMN theorem thus implies that there is a self-adjoint $A_{1,\text{int}}$, which is bounded from below.
\end{proof}

\section[Exponential decay of Schr\"odinger eigenfunctions (by Rupert L. Frank)]{Exponential decay of Schr\"odinger eigenfunctions (by Rupert L. Frank\footnote{r.frank@lmu.de; Mathe\-matisches Institut, Ludwig-Maximilans Universit\"at M\"unchen, The\-resienstr.~39, 80333 M\"unchen, Germany, and Munich Center for Quantum Science and Technology, Schel\-ling\-str.~4, 80799 M\"unchen, Germany, and Mathematics 253-37, Caltech, Pasa\-de\-na, CA 91125, USA})}\label{sec:app_b}
\renewcommand{\thefootnote}{${}$} \footnotetext{Partial support through U.S.~National Science Foundation grant DMS-1954995 and through the Deutsche Forschungsgemeinschaft (German Research Foundation) through Germany’s Excellence Strategy EXC-2111-390814868 is acknowledged.}

It is a folklore theorem that eigenfunctions of Schr\"odinger operators corresponding to eigenvalues below the bottom of their essential spectrum decay exponentially. This was raised to high art by Agmon \cite{agmon_lectures_1983} and others; see, for instance, the review \cite{Si}. It may be of interest to note that the most basic one of these bounds holds under rather minimal assumptions of the potential. This is what we record here.

Let $V\in L^1_\loc(\R^d)$ be real and set $V_\pm:=\max\{\pm V,0\}$. Given $\alpha\in[0,1]$, we say that $V_-$ is $-\Delta$-form bounded with form bound $\alpha$ if there is a $C_\alpha<\infty$ such that
$$
\int_{\R^d} V_-|\psi|^2\,dx \leq \alpha \int_{\R^d} |\nabla\psi|^2\,dx + C_\alpha \int_{\R^d} |\psi|^2\,dx 
\qquad\text{for all}\ \psi\in H^1(\R^d) \,.
$$
In this case, we define a quadratic form $h$ by
\begin{align*}
	D[h] & := \left\{ \psi\in H^1(\R^d):\ \int_{\R^d} V_+|\psi|^2\,dx <\infty \right\} \,,\\
	h[\psi] & := \int_{\R^d} \left( |\nabla\psi|^2 + V|\psi|^2\right)dx 
	\qquad\text{for}\ \psi\in D[h]	\,.
\end{align*}
This quadratic form is lower semibounded in $L^2(\R^d)$ and, if $\alpha<1$, closed. Thus, it corresponds to a selfadjoint, lower semibounded operator, which we denote by $-\Delta +V$. We abbreviate
$$
E_\infty := \inf \sigma_{\rm ess} (-\Delta+V) \in\R\cup\{+\infty\}.
$$

\begin{thm3}
	\label{expdecay}
	Assume that $V_+\in L^1_\loc(\R^d)$ and that $V_-$ is $-\Delta$-form bounded with bound $<1$. For every $E'<E_\infty$ there is a constant $C_{E'}<\infty$ such that if $E\leq E'$ and if $\psi\in D(-\Delta+V)$ satisfies $(-\Delta+V)\psi=E\psi$, then
	\begin{equation}
		\label{eq:expdecay}
		\int_{\R^d} e^{2\sqrt{E'-E}\, |x|} \left( |\nabla \psi|^2 + V_+ |\psi|^2 + (E'-E) |\psi|^2\right) dx \leq C_{E'}\, \|\psi\|^2 \,.
	\end{equation}
\end{thm3}

We emphasize that $E_\infty$ may be equal to $+\infty$, in which case $E'$ may be taken arbitrarily large. If $E_\infty<\infty$, the decay exponent $\sqrt{E'-E}$ can be any number $<\sqrt{E_\infty -E}$.

Note that under the assumptions of the theorem, $\psi$ is not necessarily bounded, so one cannot expect pointwise exponential decay bounds. The bounds in the theorem control the quantities that are natural from the definition of the operator in the form sense.

In order to prove Theorem \ref{expdecay}, we use a geometric characterization of the bottom of the essential spectrum due to Persson \cite{Per}. Let $K\subset\R^d$ be a compact set and define
$$
E_1(-\Delta+V|_{\R^d\setminus K}) = \inf\left\{ \frac{h[\psi]}{\|\psi\|^2} :\ \psi\in D[h],\, \psi\equiv 0 \ \text{on}\ K \right\}.
$$
Clearly, $E_1(-\Delta+V|_{\R^d\setminus K})$ is nondecreasing in $K$ and therefore its supremum over all compact $K\subset\R^d$ exists in $\R\cup\{+\infty\}$.

\begin{thm3}\label{perssoness}
	Assume that $V_+\in L^1_\loc(\R^d)$ and that $V_-$ is $-\Delta$-form bounded with bound $<1$. Then
	$$
	E_\infty = \sup_{K\subset \R^d\ \mathrm{compact}} E_1(-\Delta+V|_{\R^d\setminus K}) \,.
	$$
\end{thm3}

We first assume Theorem \ref{perssoness} and show how it implies Theorem \ref{expdecay}. Then we will provide a proof of Theorem \ref{perssoness} under our assumptions on $V$.

\begin{proof}[Proof of Theorem \ref{expdecay}]
	Fix $E_\infty > E''>E'$. By Theorem \ref{perssoness}, there is an $R'>0$ such that
	$$
	h[u] \geq E'' \|u\|^2
	$$
	for all $u\in D[h]$ with $u\equiv 0$ in $\overline{B_{R'/2}}$. Next, for an $R>0$ to be specified, we choose two smooth, real-valued functions $\chi_<$ and $\chi_>$ on $\R^d$ such that
	\begin{equation}
		\label{eq:perssonims1}
		\supp\chi_< \subset \overline{B_{2R}} \quad\text{and}\quad \supp\chi_> \subset \R^d\setminus B_{R}
	\end{equation}
	and such that $\chi_<^2+\chi_>^2\equiv 1$ on $\R^d$. By scaling an $R$-independent quadratic partition of unity, we may assume that
	\begin{equation}
		\label{eq:perssonims2}
		|\nabla\chi_<|^2 + |\nabla\chi_>|^2 \leq C R^{-2}
	\end{equation}
	with a constant $C$ independent of $R$. By increasing $R'$ if necessary, we can make sure that $C (R')^{-2} \leq (E''-E')/2 =:\epsilon$ with $C$ from \eqref{eq:perssonims2}. Let $f:\R^d\to\R$ be a bounded Lipschitz function and take $\varphi=e^{2f}\psi\in D[h]$ as a trial function in the quadratic form version of the equation $(-\Delta+V)\psi=E\psi$ to obtain, after an integration by parts,
	\begin{equation}
		\label{eq:agmon}
		E \int_{\R^d} e^{2f} |\psi|^2 \,dx = \int_{\R^d} \left( |\nabla(e^f \psi)|^2 + (V- |\nabla f|^2) |e^{f} \psi|^2 \right) dx \,.
	\end{equation}
	Thus, in view of the IMS formula (see, e.g., \cite[Theorem 3.2]{CyFrKiSi}),
	\begin{align*}
		E \int_{\R^d} | e^{f} \chi_< \psi|^2 \,dx + E \int_{\R^d} |e^{f} \chi_> \psi|^2 \,dx
		& = \int_{\R^d} \left( |\nabla(e^f \chi_< \psi)|^2 + \tilde V |e^f \chi_< \psi|^2 \right) dx \\
		& \quad + \int_{\R^d} \left( |\nabla(e^f \chi_> \psi)|^2 + \tilde V |e^f \chi_> \psi|^2 \right) dx
	\end{align*}
	with $\tilde V := V-|\nabla f|^2 - |\nabla \chi_<|^2- |\nabla\chi_>|^2$. For $R\geq R'$ we bound the terms on the right side from below by
	$$
	\int_{\R^d} \left( |\nabla(e^f \chi_< \psi)|^2 + \tilde V |e^f \chi_< \psi|^2 \right)dx \geq \left(E_1-\|\nabla f\|_\infty^2 -\epsilon\right) \int_{\R^d} | e^{f} \chi_< \psi|^2 \,dx
	$$
	with $E_1:=\inf\sigma(-\Delta+V)$, and
	$$
	\int_{\R^d} \left( |\nabla(e^f \chi_> \psi)|^2 + \tilde V |e^f \chi_> \psi|^2 \right)dx
	\geq \left( E'' - \|\nabla f\|_\infty^2 - \epsilon\right) \int_{\R^d} | e^{f} \chi_> \psi|^2 \,dx \,.
	$$
	Thus,
	\begin{align*}
		\left( E'' -E - \|\nabla f\|_\infty^2 - \epsilon\right) \int_{\R^d} | e^{f} \chi_> \psi|^2 \,dx
		& \leq \left(E-E_1+\|\nabla f\|_\infty^2 + \epsilon\right) \int_{\R^d} | e^{f} \chi_< \psi|^2 \,dx \,,
	\end{align*}
	and therefore
	\begin{align*}
		\left( E'' -E - \|\nabla f\|_\infty^2 - \epsilon\right) \int_{\R^d} | e^{f} \psi|^2 \,dx
		& \leq \left(E'' -E_1 \right) \int_{\R^d} | e^{f} \chi_< \psi|^2 \,dx \\
		& \leq \left(E'' -E_1 \right) \| \psi\|^2 \ \sup_{B_R} e^{2f} \,.
	\end{align*}
	Ideally, we would want to choose $f(x)=\kappa|x|$ with $\kappa$ as large as possible. The wish to have a positive constant ($\epsilon$, say) in front of the integral on the left side then dictates our choice $\kappa=\sqrt{E''-E-2\epsilon}=\sqrt{E'-E}$. The problem with this `ideal' choice of $f$ is that the function $|x|$ is Lipschitz, but not bounded. We remedy this by taking $|x|/(1+\delta|x|)$ instead and proving bounds which are uniform in the parameter $\delta>0$, which we will let tend to zero at the end. Thus, let us choose
	$$
	f(x):= \sqrt{E'-E}\ \frac{|x|}{1+\delta|x|}
	$$
	with a (small) parameter $\delta>0$. This is a Lipschitz function satisfying $\|\nabla f\|_\infty = \sqrt{E'-E}$. Thus, the previous inequality with $R=R'$ becomes
	\begin{align*}
		\epsilon \int_{\R^d} | e^{f} \psi|^2 \,dx
		& \leq \left(E'' -E_1 \right) \| \psi\|^2 \ e^{2R' \sqrt{E'-E}} \,.
	\end{align*}
	Since the right side is independent of $\delta$, we can take the limit $\delta\to 0$ and obtain by monotone convergence
	\begin{align*}
		\epsilon \int_{\R^d} | e^{\sqrt{E'-E} |x|} \psi|^2 \,dx
		& \leq \left(E''-E_1 \right) \| \psi\|^2 \ e^{2R' \sqrt{E'-E}} \,.
	\end{align*}
	This is already one of the inequalities claimed in the theorem.
	
	To prove boundedness of the terms involving the gradient term and $V_+$ we recall that, by form boundedness,
	$$
	h[e^f \psi] \geq (1-\alpha) \int_{\R^d} |\nabla (e^f\psi)|^2 \,dx + \int_{\R^d} V_+ |e^f\psi|^2 \,dx -C_\alpha \int_{\R^d} |e^f\psi|^2 \,dx \,.
	$$
	This, together with identity \eqref{eq:agmon}, implies
	$$
	\left(E + \|\nabla f\|_\infty^2 +C_\alpha \right) \int_{\R^d} |e^f\psi|^2 \,dx \geq (1-\alpha) \int_{\R^d} |\nabla (e^f\psi)|^2 \,dx + \int_{\R^d} V_+ |e^f\psi|^2 \,dx \,.
	$$
	Using
	\begin{align*}
		|\nabla (e^f\psi)|^2 & = e^{2f} |\nabla\psi + \psi \nabla f|^2 = e^{2f} \left( |\nabla\psi|^2 + 2 \re\overline\psi\nabla\psi\cdot\nabla f + |\psi|^2|\nabla f|^2 \right) \\
		& \geq e^{2f} \left( \frac 12 |\nabla\psi|^2 - |\psi|^2|\nabla f|^2 \right),
	\end{align*}
	we obtain
	$$
	\left(E + (2-\alpha) \|\nabla f\|_\infty^2 +C_\alpha \right) \int_{\R^d} |e^f\psi|^2 \,dx \geq \frac{1-\alpha}2 \int_{\R^d} |e^f \nabla \psi|^2 \,dx + \int_{\R^d} V_+ |e^f\psi|^2 \,dx \,.
	$$
	Since we have already shown an upper bound on the left side, this completes the proof of the theorem.
\end{proof}

Thus, we are left with proving Theorem \ref{perssoness}. We use the following abstract characterization of the essential spectrum.

\begin{thm2}\label{bottomess}
	Let $a$ be a lower semibounded, closed quadratic form in a Hilbert space and $A$ the corresponding self-adjoint operator. Then
	$$
	\inf \sigma_{\rm ess}(A) = \inf\left\{ \liminf_{j\to \infty} a[\xi_j] :\ \xi_j\rightharpoonup 0 \,,\ \|\psi_j\|=1 \right\}
	$$
	(with the convention that $\inf\emptyset=+\infty$). Moreover, if both sides are finite, then there is a sequence $(\xi_j)$ with $\|\xi_j\|=1$, $a[\xi_j]\to\inf \sigma_{\rm ess}(A)$ and $\xi_j\rightharpoonup 0$ in $D[a]$.
\end{thm2}

This lemma is classical. The proof in \cite[Lemma 1.20]{FrLaWe} shows the first assertion and, in the case of finiteness, the existence of a normalized sequence with $a[\xi_j]\to \inf \sigma_{\rm ess}(A)$ and $\xi_j\rightharpoonup 0$. Since this sequence is bounded in $D[a]$, a subsequence converges weakly in $D[a]$ and, since $D[a]$ is continuously embedded into the Hilbert space, the weak limit is necessarily zero, as claimed.

\begin{proof}[Proof of Theorem \ref{perssoness}]
	We abbreviate $E_\infty' := \sup_{K \ \textnormal{compact}} E_1(-\Delta+V|_{\R^d\setminus K})$.
	
	We begin by proving $E_\infty\geq E_\infty'$. We may assume that $E_\infty<\infty$ and we shall show that for all $R>0$,
	\begin{equation}
		\label{eq:perssonessproof0}
		E_1(-\Delta+V|_{B_R^c}) \leq E_\infty \,,
	\end{equation}
	for then the claimed inequality follows as $R\to\infty$. Fix $R>0$ and let $\chi_<$ and $\chi_>$ be as in the proof of Theorem \ref{expdecay}. By Lemma \ref{bottomess}, there is a sequence $(\xi_j)\subset D[h]$ with $\|\xi_j\|=1$ such that $\xi_j\rightharpoonup 0$ in $D[h]$ and $h[\xi_j]\to E_\infty$. Then
	\begin{equation}
		\label{eq:perssonessproof}
		E_1(-\Delta+V|_{B_R^c}) \leq h\left[ \frac{\chi_>\xi_j}{\|\chi_>\xi_j\|}\right]
	\end{equation}
	and our goal is to estimate the right side as $j\to\infty$.
	
	By Rellich's compactness theorem, $\xi_j\to 0$ in $L^2_\loc(\R^d)$, so $\chi_<\xi_j\to 0$ in $L^2(\R^d)$ and
	\begin{equation}
		\label{eq:perssonessproof1}
		\|\chi_>\xi_j\|^2 = \|\xi_j\|^2 - \|\chi_<\xi_j\|^2 \to 1
		\qquad\text{as}\ j\to\infty \,.
	\end{equation}
	Moreover, by the IMS formula,
	\begin{equation}
		\label{eq:perssonessproof2}
		h\left[\chi_>\xi_j\right] = h\left[\xi_j\right] - h\left[\chi_<\xi_j\right] + \left\| \left(|\nabla\chi_<|^2+ |\nabla\chi_>|^2\right)^{1/2}\xi_j\right\|^2.
	\end{equation}
	The last term vanishes as $j\to\infty$ again by Rellich's theorem. Moreover,
	$$
	h\left[\chi_<\xi_j\right] \geq E_1 \|\chi_<\xi_j\|^2
	$$
	and therefore
	$$
	\liminf_{j\to\infty} h\left[\chi_<\xi_j\right] \geq \liminf_{j\to\infty} E_1 \|\chi_<\xi_j\|^2 = 0 \,. 
	$$
	Putting this into \eqref{eq:perssonessproof2}, we learn that
	$$
	\limsup_{j\to\infty} h\left[\chi_>\xi_j\right] \leq \limsup_{j\to\infty} h\left[\xi_j\right] = E_\infty \,.
	$$
	This, together with \eqref{eq:perssonessproof} and \eqref{eq:perssonessproof1}, yields \eqref{eq:perssonessproof0}.
	
	We now prove the converse inequality $E_\infty\leq E_\infty'$. Let $(R_j)\subset(0,\infty)$ be a sequence with $R_j\to\infty$ and let $(\psi_j)\subset D[h]$ be a sequence with $\|\psi_j\|=1$, $\psi_j\equiv 0$ in $\{|x|<R_j\}$ and $h[\psi_j] - E_1(-\Delta+V|_{B^c_{R_j}})\to 0$. The support condition implies that $\psi_j\rightharpoonup 0$ in $L^2(\R^d)$ and therefore, by Lemma \ref{bottomess},
	$$
	E_\infty \leq \liminf_{j\to\infty} h[\psi_j] = \liminf_{j\to\infty} E_1(-\Delta+V|_{B^c_{R_j}}) \leq E_\infty' \,,
	$$
	which proves the theorem.
\end{proof}




	

	



\bibliographystyle{abbrv}

\end{document}